\documentclass{emulateapj}
\slugcomment{The Astrophysical Journal, 834:3 (16pp), 2017 July 1} 
\usepackage{natbib}

\usepackage{amsmath}
\usepackage{color}

\newcommand{\myemail}{tra3595@rit.edu}

\shorttitle{Modeling Dusty Torus}
\shortauthors{Almeyda et al.}

\begin{document}

\title{Modeling the infrared reverberation response of the circumnuclear dusty torus in AGN: the effects of cloud orientation and anisotropic illumination}

\author{Triana Almeyda$^1$}
\author{Andrew Robinson$^1$}
\author{Michael Richmond$^1$}
\author{Billy Vazquez $^1$}

\author{Robert Nikutta $^{2,3}$}

\affil{$^1$School of Physics and Astronomy, Rochester Institute of Technology, Rochester, NY 14623, USA; \myemail}
\affil{$^2$National Optical Astronomy Observatory, 950 N Cherry Ave, Tucson, AZ 85719, USA}
\affil{$^3$Instituto de Astrof\'isica, Facultad de F\'isica, Pontificia Universidad Cat\'olica de Chile, 306, Santiago 22, Chile}
%\affil{$^1$\textit {\{tra3595\}@rit.edu}}

%\author{Triana Almeyda\altaffilmark{1}}
%\author{Andrew Robinson\altaffilmark{1}}
%\author{Michael Richmond\altaffilmark{1}}
%\author{Billy Vazquez \altaffilmark{1}}
%
%\author{Robert Nikutta \altaffilmark{2,3}}
%\email{tra3595@rit.edu}
%
%\altaffiltext{1}{School of Physics and Astronomy, Rochester Institute of Technology, Rochester, NY 14623, USA}
%\altaffiltext{2}{National Optical Astronomy Observatory, 950 N Cherry Ave, Tucson, AZ 85719, USA}
%\altaffiltext{3}{Instituto de Astrof\'isica, Facultad de F\'isica, Pontificia Universidad Cat\'olica de Chile, 306, Santiago 22, Chile}
%

\begin{abstract}

The obscuring circumnuclear torus of dusty molecular gas is one of the major components of active galactic nuclei (AGN). 
The torus can be studied by analyzing the time response of its infrared (IR) dust emission to variations in the AGN continuum luminosity, a technique known as reverberation mapping. The IR response is the convolution of the AGN ultraviolet/optical light curve with a transfer function that contains information about the size, geometry, and structure of the torus. Here, we describe a new computer model that simulates the reverberation response of a clumpy torus. Given an input optical light curve, the code computes the emission of a 3D ensemble of dust clouds as a function of time at selected IR wavelengths, taking into account light travel delays. We present simulated dust emission responses at 3.6, 4.5, and 30 $\mu$m that explore the effects of various geometrical and structural properties, dust cloud orientation, and anisotropy of the illuminating radiation field. 
We also briefly explore the effects of cloud shadowing (clouds are shielded from the AGN continuum source). Example synthetic light curves have also been generated, using the observed optical light curve of the Seyfert 1 galaxy NGC 6418 as the input. The torus response is strongly wavelength-dependent, due to the gradient in cloud surface temperature within the torus, and because the cloud emission is strongly anisotropic at shorter wavelengths. Anisotropic illumination of the torus also significantly modifies the torus response, reducing the lag between the IR and optical variations. 

\end{abstract}

\keywords{ dust, extinction--- galaxies: active --- galaxies: nuclei --- galaxies: Seyfert --- infrared: galaxies--- radiative transfer} --- %accretion disks ---galaxies: individual (NGC 6418)}

\section{Introduction}

Over the last few decades, it has been established that supermassive black holes (SMBH) are present in the centers of most, if not all, galaxies \citep{Ferrarese:2005aa,Fabian:2012aa}. During growth phases, these SMBH are observed as active galactic nuclei (AGN); the SMBH grows by accreting interstellar gas while releasing vast amounts of energy in the form of electromagnetic radiation, gas outflows, and jets of ionized plasma. However, active SMBH are surrounded by dusty molecular gas, which can obscure our observations and severely hinder studies of the SMBH growth phase \citep{Ricci:2017aa}. For this reason, it is important to understand the properties (such as size, structure, and composition) of these obscuring structures. 

AGN were originally classified based on their spectroscopic signatures, ranging from Type 1 AGN exhibiting both broad and narrow emission lines to Type 2 AGN exhibiting only narrow lines. However, after observing broad emission line features in scattered light in a Type 2 AGN, \cite{Antonucci:1985aa} proposed that Type 2s are, in fact, Type 1 AGN that are obscured in our line of sight by a ``very thick absorbing disk". This observation led to the development of unified models of AGN where the central engine and broad emission line regions (BLRs) are surrounded by a dusty region believed to be roughly in the form of a torus \citep{Krolik:1986aa,Krolik:1988aa,Antonucci:1993aa,Urry:1995aa}. 

Current single-dish telescopes do not have the spatial resolution to directly image the torus, so it is necessary to use other methods to study its size and structure. 
Radiative transfer models of the dusty torus have been developed to reproduce the observed infrared (IR) spectral energy distributions (SEDs) of AGN. The main constraints to which every torus model must conform are: the torus is dusty, has a high optical depth in the optical/ultraviolet (UV), is geometrically thick, and has a size scale of a few parsecs \citep{Hoenig:2013aa}. At first, the torus was modeled as a smooth density distribution \citep{Pier:1992aa,Granato:1994aa,Efstathiou:1995aa,Fritz:2006aa,Feltre:2012aa,Efstathiou:2013aa,Efstathiou:2014aa} to simplify the radiative transfer problem. 
However, \citealp{Krolik:1988aa} (and others since) have argued that, theoretically, the torus should be clumpy; this is also supported by observational evidence both in the IR (e.g., \citealp{Shi:2006aa},\citealp{Tristram:2007aa, Burtscher:2009aa}) and X-rays (\citealp{Markowitz:2014aa} and references therein).

Several different clumpy torus models have been developed (for example, \citealp{Nenkova:2002aa,Dullemond:2005aa, Honig:2006aa, Nenkova:2008aa,Nenkova:2008ab,Schartmann:2008aa, Honig:2010aa, Stalevski:2012aa}). These models have produced IR SEDs that are, generally, in close agreement with observations. They also successfully reproduce the differences in strength of the 10 $\mu$m silicate features, as seen in Type 1 and Type 2 AGN \citep{Nenkova:2008ab,Nikutta:2009aa}. High-resolution mid-IR (MIR) imaging and interferometric observations have provided evidence that the torus is compact, with a size of a few pc \citep{Jaffe:2004aa,Poncelet:2006aa,Packham:2007aa,Tristram:2007aa,Raban:2009aa,Burtscher:2013aa, Lopez-Gonzaga:2014aa,Tristram:2014aa}. Recent Atacama Large Millimeter/submillimeter Array (ALMA) observations have revealed a disk-like structure in NGC 1068 (one of the closest AGN) within 10 pc, which may be the sub-millimeter counterpart of the torus \citep{Gallimore:2016aa,Garcia-Burillo:2016aa,Imanishi:2016aa}.
Clumpy models can reproduce the IR SED with a compact torus, consistent with the observed sizes.
In contrast, the SED fits produced by smooth distribution models require tori extending beyond 100s of pc.

At present, the only way to resolve the dust distribution in the inner nuclear regions of AGN (nearby ones, at least) is through IR interferometry. Near-IR (NIR) interferometry observations have provided estimates for the innermost torus dust radius for several objects \citep{Swain:2003aa,Kishimoto:2009aa,Pott:2010aa,Kishimoto:2011aa,Weigelt:2012aa}. 

However, IR interferometry can only resolve the torus in bright, relatively nearby AGN. The current state of this technique still falls short of delivering true images and interpretation of the data requires comparison of an image model with the observed correlated fluxes. However, the forthcoming Multi-AperTure mid-Infrared SpectroScopic Experiment (MATISSE) instrument is expected to be capable of MIR closure-phase aperture-synthesis imaging of several of the nearest AGN \citep{Lopez:2014aa}.

A complementary method that relies on fine time sampling, not spatial resolution, and thus can probe galaxies at higher redshifts is reverberation mapping. This technique was originally applied to the BLR, and has been extensively developed in that context (e.g., \citealp{Blandford:1982aa,Peterson:1993aa, Peterson:2014aa,Shen:2016aa} and references therein). It can also be applied to the torus, as the dust absorbs the optical/UV light emitted by the accretion disk and re-emits it in the IR. The dust emission will, therefore, vary in response to driving variations in the AGN optical/UV luminosity. Due to differences in light travel times to different points in the torus, the IR radiation from a particular dusty clump will reach the observer at a delayed time that depends on the orientation and position of the clump within the torus. At a given time delay, the observer sees the change in the dust emission from the part of the torus that intersects a corresponding isodelay surface \citep{Peterson:2001aa}.  The delayed response of the torus IR emission thus yields information about its size and structure. In particular, the lag between the UV/optical and IR variations is $\tau_{\rm lag} \sim r_{\rm dust}/c$, where $r_{\rm dust}$ is the dust reverberation radius, and $c$ is the speed of light. Thus, the lag provides an estimate of the size of the dust emitting region.

In a pioneering study, \cite{Clavel:1989aa} analyzed UV, optical, and IR light curves for Fairall 9 and found that the UV and optical variations were nearly identical, but the IR light curves lagged the UV and optical, with the lag increasing with IR wavelength. The results of \cite{Clavel:1989aa} motivated \cite{Barvainis:1992aa} to create a clumpy dust reverberation model that was able to reproduce the NIR light curves of Fairall 9. These early reverberation studies show that the NIR light curves did indeed follow variations in the UV/optical continuum, with lags consistent with the idea that the NIR emission is from the dust that surrounds the central source. 

The dust reverberation radius has now been measured by monitoring the optical (V-band) and NIR (K-band) emission for about 20 AGN (e.g., \citealp{Nelson:1996aa,Oknyanskij:2001aa,Minezaki:2004aa,Suganuma:2006aa,Koshida:2014aa,Pozo-Nunez:2014aa}). Dust emission in the K-band is believed to represent the innermost torus radius, because the temperature of dust emitting at this wavelength is close to the dust sublimation temperature. These reverberation studies have verified that the torus inner radius is indeed larger than that determined for the BLR. Furthermore, the radii determined by dust reverberation correlate tightly with AGN luminosity, $L_{\rm AGN}$, as $r\propto {L_{\rm AGN}}^{0.5}$ (\citealp{Suganuma:2006aa, Koshida:2014aa}). 
Radii determined from NIR interferometry follow a similar $R-L$ relationship and are of the same order as (or slightly larger than) radii determined using reverberation mapping methods \citep{Kishimoto:2009aa,Kishimoto:2011aa, Burtscher:2013aa, Koshida:2014aa}.  
However, the radii determined by these methods are systematically smaller than those based on the theoretical dust sublimation radius predicted for a sublimation temperature of 1500 K for an ISM dust composition and grain size of 0.05 $\mu$m \citep{Kishimoto:2007aa}. 

Various explanations have been proposed to explain this discrepancy. In order to model the NIR bump seen in Type 1 AGN SEDs, a hot graphite dust component, represented as a blackbody spectrum, typically has to be added \citep{Mor:2009aa,Mor:2012aa}. Graphite dust has a higher sublimation temperature than silicate dust and thus can survive closer to the accretion disk. Although this can produce reasonable agreement with the short observed lags, torus radiative transfer models that include multi-grain components do not produce the observed NIR bump (\citealp{Schartmann:2008aa} and references within).  
Alternatively, \cite{Kishimoto:2007aa} suggested that larger grain sizes can explain the smaller observed radii, because the dust sublimation radius varies with grain size as $a^{-1/2},$ where $a$ is the size of the grain. 
Another possibility is anisotropic illumination of the torus. \cite{Kawaguchi:2010aa,Kawaguchi:2011aa} considered a torus reverberation model that included the effects of edge darkening of the accretion disk, as well as torus self-occultation and misalignment between the torus and the accretion disk axes. They found that it produced lags comparable with the observations.

Most of the authors of this paper are members of a large international collaboration that is engaged in the first attempt to reverberation map the torus at mid-IR wavelengths. This collaboration conducted a 2.5 year monitoring campaign, during which 12 broad-line AGN were observed using the \textit{Spitzer Space Telescope} at 3.6 and 4.5 $\mu$m, supported by ground-based optical observations. 
The goal is to use these data to determine the sizes and structures of the tori for these AGN. The observed reverberation radii for one object, NGC 6418, have been published for the first half of the campaign \citep{Vazquez:2015aa} and a second paper reporting the reverberation radii for the entire campaign is currently in preparation. Reverberation analysis of the optical-IR light curves of the remaining objects is ongoing; \cite{Vazquez:2015ab} presents the Spitzer light curves for the whole sample. 

The measured optical-IR lags only give an indication of the size of the torus.
However, the reverberation response is also sensitive to the geometry and orientation of the dust distribution, as well as physical processes such as dust sublimation and reformation. Thus, suitable models, mapping the dust emission response as a function of wavelength, are needed to interpret the lags and, ideally, extract additional structural information contained in the data. 

The aim of this work is to introduce a new reverberation mapping code that models the IR response within the clumpy torus paradigm. Using an observed AGN optical light curve as the ``input" signal, this code computes the torus response for a variety of structural parameters, such as torus inclination, cloud distribution, and radial extent, for direct comparison to the observed IR light curves. Due to the large parameter space, it is useful to compute a library of approximate transfer functions to explore how the various parameters affect the dust emission response.
These results will be presented in two papers. The focus of this first paper is to explore the effects of (1) dust cloud orientation (the cloud's dust emission is anisotropic because the illuminated surface is hotter) and (2) anisotropic illumination of the torus by the AGN radiation field. 
The second paper will provide a comprehensive exploration of the effects of self-occultation, and the structural and geometrical parameters of the torus.
This paper is organized as follows. The computer model will be outlined in Section 2. We will present examples of computed approximate transfer functions in Section 3 followed by a discussion of the implications of our results in Section 4. Lastly, in Section 5, we will conclude with a summary of our findings and an outline of the future work we plan to implement to enhance our model.

\section{Outline of Model: TORMAC}\label{sec:model}

%A computer code, TORMAC (TOrus Reverberation MApping Code), has been
The TOrus Reverberation MApping Code (TORMAC) is a computer code developed to study the time response of the thermal dust emission to variations in the AGN UV/optical continuum with respect to the key parameters that influence the response. TORMAC is based on the response mapping code of Robinson and collaborators (e.g., \citealp{Robinson:1990aa, Perez:1992aa}) and adopts a  geometry similar to that used in the CLUMPY torus model described in \cite{Nenkova:2008ab}. 

	\subsection{Torus Geometry and Structure}\label{sec:structure}
	
\begin{figure}[t]
\begin{center}
	\includegraphics[scale=0.45]{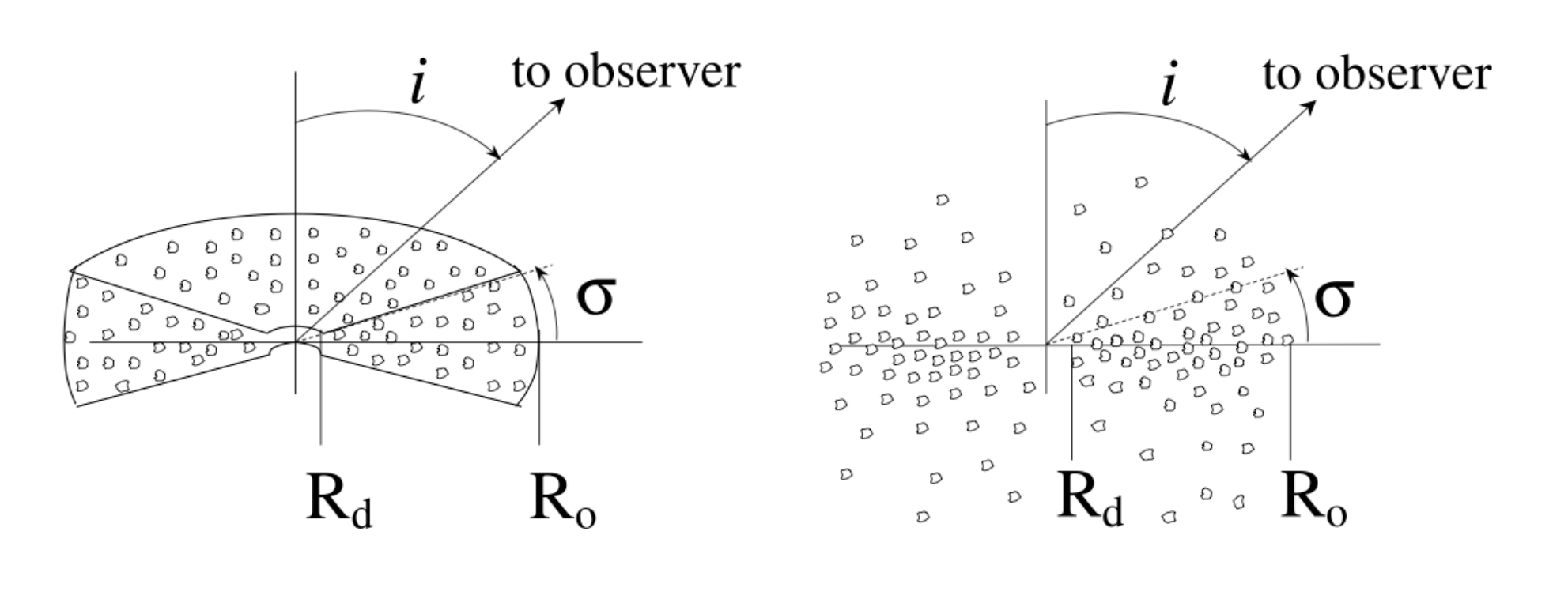}
\end{center}
\caption{Different torus geometries explored in TORMAC. The image on the left illustrates the torus geometry, with a sharp edge as defined by the angular width, $\sigma$. The image on the right illustrates the torus geometry with a fuzzy edge (clouds are distributed using a Gaussian above/below the equatorial plane). (Reproduced with permission from \citealp{Nenkova:2008ab}).}%; DOI:10.1086/590483.)}
\label{torusedge}
\end{figure}

The torus is treated as a 3D ensemble of clouds within a flared disk, centered on the AGN (Figure~\ref{torusedge}). 
The clouds are randomly distributed in spherical polar coordinates ($r$, $\beta$, $\phi$), where $r$ is the radial distance from the central source, $\beta$ is the angle measured from the equatorial plane (i.e., the complement of the polar angle, $\beta = 90^{\circ} - \theta$), and $\phi$ is the azimuthal angle. In the radial direction, the clouds are distributed
following a power law with index $p$. In $\phi$, they are distributed uniformly; in $\beta$, they follow either a uniform or Gaussian distribution.

The height of the torus above the mid-plane is defined by the angular width, $\sigma$; and the surface boundary can be defined to be ``sharp", meaning that the clouds are distributed uniformly in $\cos\beta$, but confined within the angular width, $|\beta|\le\sigma$, or ``fuzzy", meaning that the clouds are distributed in $\beta$ according to a Gaussian with width $\sigma$, and centered on the equator. The angular width, $\sigma$, can range from small values corresponding to a thin disk to $\sigma$=90$^{\circ}$, corresponding to a ``sphere" when the edge is sharp. These two different edge configurations are shown in Figure~\ref{torusedge}. The inclination angle, $i$, of the torus ranges from 0$^{\circ}$ (face-on) to 90$^{\circ}$ (edge-on) with respect to the observer.

In the radial direction, the number of clouds between $r$ and $r+dr$ from the central source is represented as $dN(r)=N(r)dr=A_{\rm o}(r/R_{\rm o})^{p}dr \propto r^{p},$ where $A_{\rm o}$ is the normalization constant and $R_{\rm o}$ is the outer radius of the torus. The normalization constant, $A_{\rm o}$, is set by the condition $\int_{R_{\rm d}}^{R_{\rm o}}  A_{\rm o}(r/R_{\rm o})^{p} dr = N$, where $N$ is a free parameter that represents the total number of clouds in the torus.  The cloud number density is given by $n(r)=dN(r)f(\sigma) /(4\pi r^{2} \sin\sigma dr)$, where $f(\sigma) = 1$ when $|\beta|\le\sigma$ for the ``sharp" case, or $f(\sigma) = \exp[-\beta^2/\sigma^2]$ for the ``fuzzy'' case. The clouds are treated as optically thick points with an associated surface area, i.e., we assume that the cloud's size, $R_{\rm cl} << r$.

The radial extent of the torus is defined by $Y=R_{\rm o}/R_{\rm i}$, where $Y$ is a free parameter and $R_{\rm i}$ is the inner radius. In the current version of TORMAC, the inner radius of the torus is taken to be the closest distance to the AGN at which dust can survive without being destroyed by sublimation. This radius, which is the dust sublimation radius ($R_{\rm d}\equiv R_{\rm i}$) assuming an ISM grain mixture, is given by
\begin{equation}\label{dustsub}
R_{\rm d}\simeq 0.4\left(\frac{L_{\rm AGN}}{10^{45} \,\rm erg\, s^{-1}}\right)^{1/2}\left(\frac{1500\,\rm K}{T_{\rm sub}}\right)^{2.6} \rm pc,
\end{equation}
where $T_{\rm sub}$ is the dust sublimation temperature and $L_{\rm AGN}$ is the bolometric luminosity of the AGN (e.g., \citealp{Nenkova:2008ab}).

	\subsection{Cloud Shadowing}\label{sec:clshadow}
Given the optically thick nature of the clumpy torus within the AGN unified model, there is a high probability that a given line of sight from the central source passing through the torus is intercepted by at least one optically thick cloud. 
Thus, there is a high probability that a given cloud at radius $r$ within the torus will have its ``view'' of the central AGN continuum source obstructed or ``shadowed" by one or more clouds at smaller radii (i.e., closer to the central source). For instance, \citet{Nenkova:2008ab} inferred that clumpy torus models require 5--15 clouds along radial equatorial rays to explain AGN IR SEDs. The shadowed clouds are not directly heated by the UV/optical continuum. Rather, they are immersed in the diffuse radiation field produced by the surrounding directly heated clouds. The probability that a cloud at radius $r$, along a ray at angle $\beta$ from the equatorial plane, has an unobstructed view of the central continuum source is given by $P(r,\beta) \sim\exp{[-\mathcal{N}(r,\beta)]}$, where $\mathcal{N}(r,\beta)$ is the average number of interior clouds along the ray \citep{Nenkova:2008aa}. The latter quantity is given by

\begin{equation}\label{Nclos}
\mathcal{N}(r,\beta) = \int_{R_{\rm d}(\beta)}^{r} n(r') A_{\rm cl}\, dr',
\end{equation}
where $A_{\rm cl}$ is the cloud cross-sectional area, $A_{\rm cl} = \pi R_{\rm cl}^2$, which we currently assume to be constant. Note that the dust sublimation radius can be a function of $\beta$ if the torus is illuminated anisotropically, as would be expected if the continuum source is a geometrically thin accretion disk (see below in Section~\ref{sec:dustem}). The cloud radius is defined in terms of the average torus volume filling factor, $\Phi = NV_{\rm cl}/V_{\rm tor}$, where $V_{\rm cl}$ and $V_{\rm tor}$ are the volumes of an individual cloud and the torus, respectively. Thus, assuming spherical clouds, $R_{\rm cl} = [3\Phi V_{\rm tor}/4\pi N]^{1/3}$. 

In any given volume element, therefore, the dust emission is determined by the sum of contributions from $n(r)P(r,\beta)$ directly heated and $n(r)(1-P(r,\beta))$ indirectly heated (shadowed) clouds.

\subsection{Calculating the Time Dependent Dust Emission}\label{sec:timedep}

\begin{figure}[t!]
%\begin{center}
\centering
	\includegraphics[scale=.35]{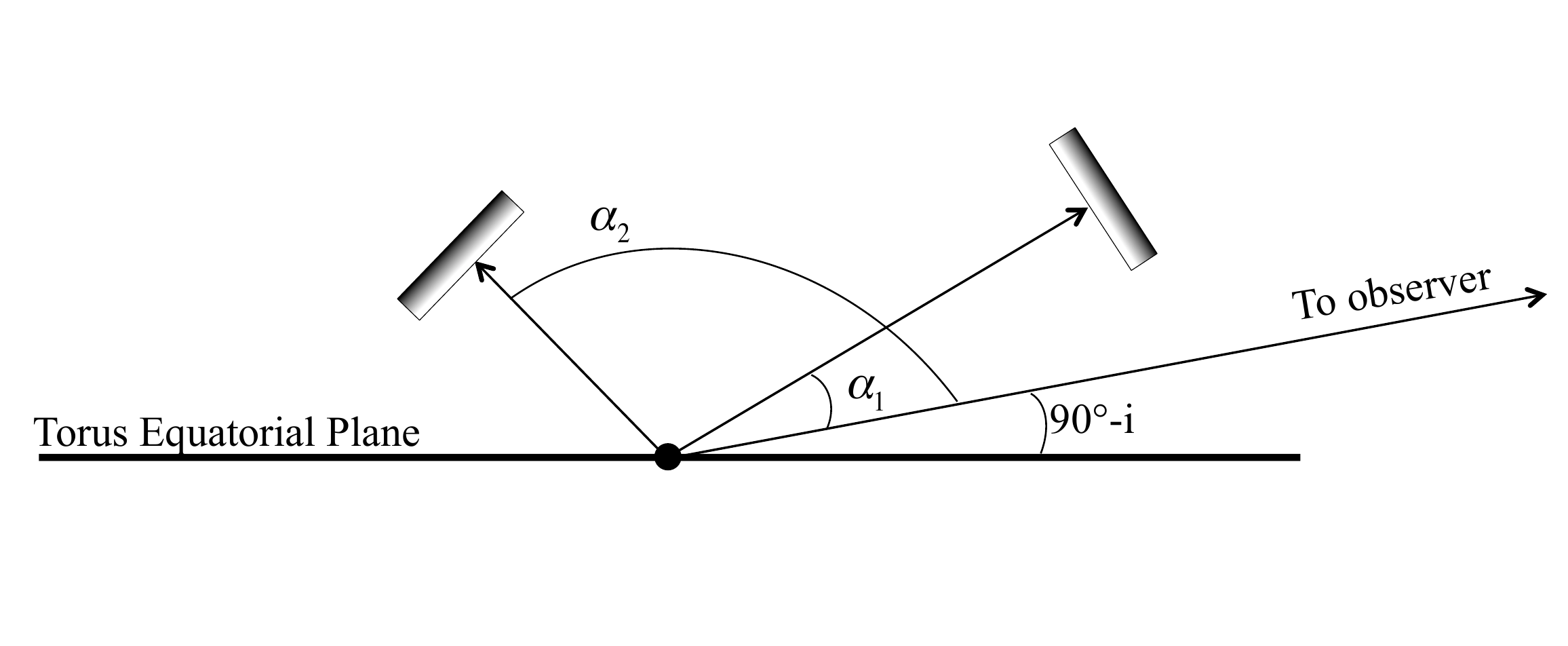}
%\end{center}
\caption{Illustration of dust cloud orientation with respect to a distant observer. The cloud is depicted as a slab here, for simplicity. The side of the slab facing the observer is illuminated and at a higher temperature than its non-illuminated side. The orientation of the slab and the side facing the observer is determined by $\alpha$, which is the angle between the source--slab and the source--observer lines of sight.}
\label{clorientdiagram}
\end{figure}

Given an input light curve, the emission spectrum of each cloud is calculated as a function of time as measured by a distant observer, depending on its location with respect to the isodelay surface corresponding to the current observer time step. In other words, at time $t$, the distant observer sees the emission from all the clouds in the torus. However, the emission of an individual cloud at the current source time $t_{\rm s}$ (i.e., the time at which the light is currently being emitted by the AGN) is actually responding to the light that was emitted by the source at an earlier time $t'_{\rm s}=t_{\rm s}-(r/c)$. 
The observer will see the cloud response after a time delay relative to the optical continuum $t=t_{\rm s}-(r/c)\cos\alpha$, where $\alpha$ is the angle between the observer's line of sight to the central source and the cloud's line of sight to the central source (see Figure~\ref{clorientdiagram}). Thus, at every observer time step, the luminosity of each cloud is calculated based on the luminosity of the AGN at time $t'_{\rm s}=t-(r/c)(1-\cos\alpha)$. Note that we are assuming that individual clouds respond on timescales that are much less than the light crossing time of the torus inner radius.

The power radiated by individual clouds is then summed to determine the total emission from the distribution of clouds within the torus at a given time. Thus, TORMAC computes the specific luminosity of the torus as a function of observer time at any wavelength. 
Hence, it is possible to investigate the shape of the torus' multi-wavelength transfer function (response map) with respect to its inclination to the observer's viewpoint and various structural parameters, including the radial and vertical thickness, and the spatial distributions of clouds within the torus.  

	\subsection{Dust Emission}\label{sec:dustem} 
The dust clouds within the torus are either heated directly by the UV/optical continuum emitted from the accretion disk, or indirectly by the diffuse radiation field produced by the directly illuminated clouds.

In TORMAC, the illuminating radiation field can be chosen to be either isotropic or anisotropic, to approximate emission from an optically thick, geometrically thin accretion disk. 
In the isotropic case, the flux incident on a directly illuminated cloud at radius $r$ is simply $F_{\rm AGN}=L/4\pi r^2$. Thus, the flux absorbed (and reprocessed) by each cloud is dependent only on its distance from the AGN. In the anisotropic case, the incident flux is also a function of polar angle; the accretion disk emits less radiation in the equatorial plane than along the disk axis. We model this ``edge darkening" effect following the prescription of \cite{Netzer:1987aa},  $F_{\rm AGN}\propto (1/3)\cos\theta(1+2\cos\theta)$, where $\theta=90^{\circ}-\beta$ is the polar angle. The sublimation radius, in this case, is a function of both polar angle and distance, 

\begin{equation}\label{Rdtheta}
R_{\rm d}(\theta)\simeq 0.4\left(\frac{L(\theta)}{10^{45} \,\rm erg\, s^{-1}}\right)^{1/2}\left(\frac{1500\,\rm K}{T_{\rm sub}}\right)^{2.6} \rm pc,
\end{equation}
where $L(\theta)=[s+(1-s)(1/3)\cos\theta(1+2\cos\theta)]L_{\rm AGN}$, and $s<1$ is a ``softening'' factor, which we introduce to prevent $R_{\rm d}(\theta)$ going to zero at $\theta =90^{\circ}$. Clouds closer to the equatorial plane (i.e., at larger $\theta$ or smaller $\beta$ values) can reside at smaller radii than is possible for isotropic illumination. For the simulations presented in this paper, $s=0.1$.

 			\subsubsection{Directly Heated Clouds}\label{sec:direct}
The dust emission from each directly illuminated cloud is determined by using a grid of dust cloud models that was created for the clumpy torus model described by \cite{Nenkova:2008aa,Nenkova:2008ab}.
This grid was constructed from radiative transfer calculations for a plane-parallel slab of dust, performed with the 1D DUSTY code of \cite{Ivezic:1999aa}. 
In these models, the dust composition is a standard Galactic mix of 53\% silicates and 47\% graphites. The dust grain size is defined by the standard Mathis-Rumpl-Nordsieck (\citealp{Mathis:1977aa}) power-law distribution $n(a)\propto a^{-3.5}$, with grain size $a$ varying from 0.005 to 0.25 $\mu$m. The dust slab, with optical depth at visual wavelengths $\tau_V$ is illuminated by the AGN continuum, which has a spectral shape represented by a broken power law \citep[][see Equation 13]{Nenkova:2008aa}. In order to approximate the emission of an arbitrarily shaped cloud, 
the model grid consists of dust emission spectra of ``synthetic clouds'', constructed by averaging the slab emission computed with DUSTY over all possible orientations with respect to the central source \citep{Nenkova:2008aa}.

The emission spectrum of a cloud within the model grid is determined by the parameters $T_{\rm cl}$, $\alpha$, and $\tau_{V}$. The cloud model grid spans a range in illuminated surface temperature of $100 \le T_{\rm cl} \le 1500$ K, optical depth of $5 \le \tau_{V} \le  150$, and cloud orientation $\alpha=$0$^{\circ}$(non-illuminated face toward observer) to 180$^{\circ}$(illuminated face toward observer). 
In the simulations presented in this paper, all clouds are considered to have the same size, with an optical depth $\tau_V = 40$. This value falls within the rather wide range ($20 \lesssim \tau_V \lesssim 150$) inferred from fits to the IR SEDs of AGN \citep{Nenkova:2008ab, Asensio-Ramos:2009aa, Mor:2009aa, Ramos-Almeida:2009aa, Honig:2010aa, Ramos-Almeida:2011ab, Alonso-Herrero:2014aa}; in clumpy torus models, the SED is not particularly sensitive to $\tau_V$ for  values $\gtrsim 20$ \citep{Nenkova:2008ab}. 
The flux emitted by a given cloud is, therefore, determined by a two-dimensional linear interpolation in cloud illuminated surface temperature and cloud orientation, $\lambda F_{\lambda}(T_{\rm cl}, \alpha)$.

The temperature of the illuminated surface of each dust cloud is determined by scaling the temperature at the inner radius (i.e., $T_{\rm sub}$, given by Equation~\ref{dustsub}) according to the incident flux based on the cloud's position and retarded time, 
\begin{equation}\label{Tcloud}
T_{\rm cl}(r,\theta,t'_{\rm s}) =T_{\rm sub}\left[\left(\frac{R_{\rm d}(\theta)}{r}\right)\left(\frac{L(t'_{\rm s})}{L_{\rm AGN}}\right)^{1/2}\right]^{1/2.6},
\end{equation}
where $L(t'_{\rm s})$ is the AGN luminosity at the retarded time $t'_{\rm s}$ (see Section~\ref{sec:timedep}) and $L_{\rm AGN}$ is the reference AGN luminosity that was used to determine the sublimation radius. For  $Y\lesssim 10$, Equation~\ref{Tcloud} gives essentially the same results as the analytic approximations derived from Figure 7 of \citep{Nenkova:2008aa}.

As the clouds are heated directly by the AGN, the side facing the AGN is illuminated; thus, it has a higher temperature than the non-illuminated side of the cloud. Therefore, the cloud emits anisotropically. Depending on where it is with respect to the observer, the emission of the cloud will be that of its illuminated or non-illuminated side. 
 Thus, the observed output emission is dependent on both the distance to the AGN and the angle $\alpha$ between the direction of the lines of sight of the cloud and the observer from the central source, see Figure~\ref{clorientdiagram}.  

The angle $\alpha$, calculated for each cloud, is a function of the cloud's position in the torus relative to the observer. Using the angular coordinates $\beta$ and $\phi$, and the torus inclination $i,$
$$\cos\alpha=\cos\beta \cos\phi \sin i+\sin\beta \cos i.$$

\begin{figure}[t!]
\begin{center}
	\includegraphics[scale=0.6]{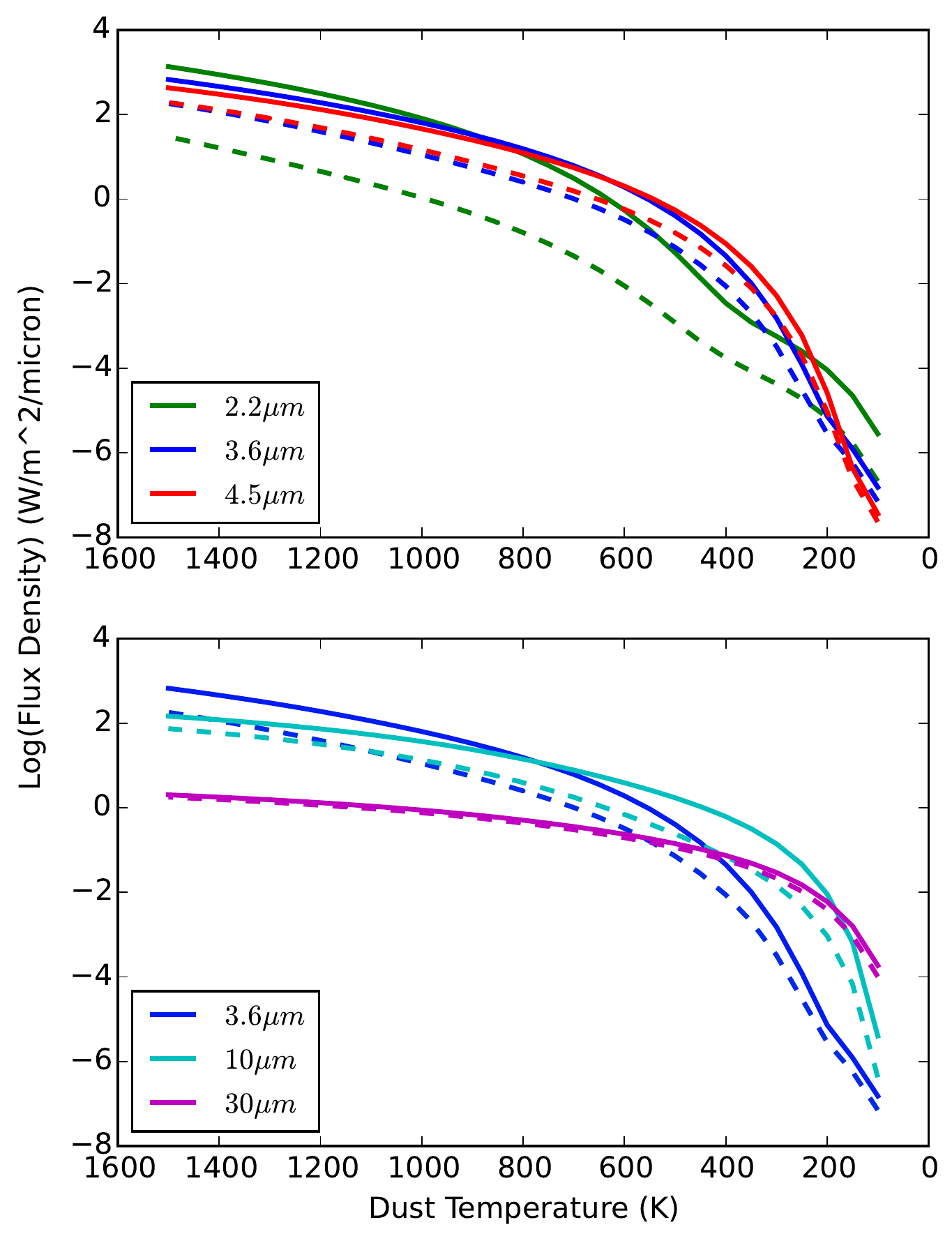}
\end{center}
\caption{The specific flux, obtained from the pre-computed dust emissivity grid, of a dust cloud at selected wavelengths as a function of illuminated surface temperature, $T_{\rm cl}$. The solid lines represent the illuminated side ($\alpha=$180$^{\circ}$) of the cloud, and the dashed lines represent the non-illuminated side ($\alpha=$0$^{\circ}$).}
\label{dusttemp}
\end{figure}

To illustrate the effect of cloud orientation, Figure~\ref{dusttemp} shows the dust emission from both the illuminated and non-illuminated sides of a cloud at 2.2, 3.6, 4.5, 10, and 30 $\mu$m as a function of the temperature of its illuminated surface. The difference between the illuminated and non-illuminated sides is highly wavelength-dependent. For example, at 2.2 $\mu$m, the flux density emitted by the illuminated surface is a factor of $\sim 100$ greater than that from the non-illuminated side at $T_{\rm cl} = 1500$ K, with this ratio decreasing gradually as $T_{\rm cl}$ decreases. In contrast, the illuminated and non-illuminated side's flux densities at 30 $\mu$m are almost the same throughout the temperature range. 
The effect on the observed emission from a given cloud is analogous to lunar phase; $\alpha$ determines the fraction of the illuminated surface of the cloud that is visible to the observer. Thus, at $\alpha$=180$^{\circ}$, the entire illuminated surface is visible, whereas at $\alpha$=0$^{\circ}$, only the non-illuminated side is visible. We henceforth refer to this as the ``cloud orientation effect".

		\subsubsection{Indirectly Heated Clouds}\label{sec:indirect}
Clouds that are ``shadowed'', i.e., whose line of sight to the central source is blocked by clouds at smaller radii, are heated ``indirectly'' by the diffuse radiation emitted by surrounding directly illuminated clouds. The dust emission in this case is also determined by interpolation in a grid of radiative transfer models created for the clumpy torus model of \cite{Nenkova:2008aa,Nenkova:2008ab}. As the heating radiation field is produced by directly heated clouds at the same location, and assumed to be isotropic, the dust emission due to indirect heating is a function only of the illuminated surface temperature ($T_{\rm cl}$) of the local directly heated clouds. In practice, the heating radiation field for a given value of $T_{\rm cl}$ (equivalent to distance from the AGN, for constant $L$) was approximated by angle-averaging the spectra of all directly heated clouds at that value of $T_{\rm cl}$ over the full range in $\alpha$ \citep{Nenkova:2008aa}. Therefore, the dust emission due to indirectly heated clouds at any given wavelength is determined by a linear interpolation in $T_{\rm cl}$, using the value determined in Equation~\ref{Tcloud}.

As pointed out by \cite{Nenkova:2008aa}, this approximation overestimates the diffuse radiation field (and therefore, the contribution to the total dust emission from indirectly heated clouds) because, in reality, a given indirectly heated cloud will see only the non-illuminated sides of the directly heated clouds in the direction of the AGN. We are also implicitly assuming that the light travel times between clouds are negligible compared with $r/c$. This is justified in the context of the local approximation for the diffuse radiation field.

	\subsection{Treatment of Dust Sublimation}\label{sec:sublimation}
As the AGN UV/optical continuum varies in brightness, the innermost dust clouds may reach temperatures greater than the dust sublimation temperature, causing the dust sublimation radius to recede as the AGN brightens. Using NIR interferometric data, \cite{Kishimoto:2013aa} studied five galaxies at multiple epochs and found evidence for a receding dust sublimation radius in one case, NGC 4151. Combining these data with previous reverberation mapping observations for this object, \cite{Kishimoto:2013aa} inferred that the dust was not completely destroyed, but eventually reformed during the phase when the AGN entered a low-luminosity state. 

The time-dependent destruction and reformation of dust clumps is not currently modeled in detail in TORMAC. Two limiting cases of dust sublimation have been implemented in the current version: in the first, the cloud surface temperature remains at the sublimation temperature and the cloud is not destroyed; in the second, the cloud is instantaneously destroyed once it reaches the sublimation temperature, and never reforms. The models presented here assume the first case where the dust cloud temperature is constrained to a maximum of $T_{\rm cl}\sim$1500 K, even when the AGN continuum luminosity is such that the computed cloud surface temperature exceeds the sublimation temperature. 

An important consequence of this assumption is that, because dust clouds are neither destroyed nor allowed to exceed $T_{\rm sub}$, the IR torus response tends to saturate. However, this prescription is preferred because, in reality, dust clouds have a large optical depth and exhibit an internal temperature gradient. Thus, 
only the dust grains on the illuminated surface of the cloud are at $T_{\rm sub}$, and the cloud will be gradually eroded rather than being completely destroyed.
The characteristic grain survival time is highly dependent on temperature and grain size \citep{Waxman:2000aa,Lu:2016aa}. The time it will take for a typical grain with a size of 0.1 $\mu$m at 1500 K to sublimate is about 1000 days. The width of the pulse ($\sim10$ days) used for the simulations presented in the next section is much less than this grain sublimation timescale, so the assumption that clouds survive is valid for this work. These issues will be addressed in more detail in future papers.

\section{Torus Dust Response Functions}\label{sec:resfuncs}
This section presents the dust emission response computed using the model described above. We have taken a step-by-step approach to exploring the manifold factors that can influence the torus response. In this paper, we focus on the effects of cloud orientation and anisotropic illumination of the torus, which were identified as candidates for explaining certain features evident in the IR light curves produced by the Spitzer monitoring campaign (for example, short optical-IR lags). Accordingly, the models presented in Sections~\ref{sec:spherical} and~\ref{sec:disk} below include only directly heated clouds. The effects of cloud shadowing are not included in the models presented in this section; instead, we defer their discussion to Section~\ref{sec:shadowing}. We also note that this version of TORMAC assumes that the radiation produced by each cloud is able to escape the torus, i.e., self-occultation is not included. A comprehensive exploration of the effects of self-occultation, as well as  the effects of effects of geometrical and structural parameters, will be presented in the second paper.

The response of the IR emission to the driving time variability of the AGN UV/optical continuum can be expressed as the convolution of the UV/optical light curve with a transfer function, whose form depends on the geometrical and structural properties of the torus, 
\begin{equation}\label{eq:transeq}
L(t)=\int_{-\infty}^{\infty} \! \Psi(\tau')L_{\rm c}(t-\tau') d\tau',
\end{equation}
where $\tau'$ is an arbitrary delay,  $\Psi(\tau')$ is the transfer function, $L(t)$ is the IR light curve, and $L_{\rm c}(t-\tau')$ is the continuum light curve at an earlier time \citep{Peterson:1993aa}. 
The transfer function is considered to be the response to a delta function input continuum pulse. In practice, we represent the input pulse as a square wave with a finite amplitude and a duration that is much less than the light-crossing time of the torus inner radius. Thus, the simulations presented here are not exact transfer functions, but close approximations that we hereafter refer to as ``response functions''.

The parameter values chosen for the models presented here are as follows. The radial extent of the torus, represented by the ratio between the outer and inner torus radius, was set to either $Y=2$ or $\textbf{10}$ representing, respectively, a compact or extended torus. Three values of the inclination of the torus axis to the observer's line of sight were also considered: $i$=\textbf{0$^{\circ}$} (face-on),  45$^{\circ}$, or $90^{\circ}$ (edge-on). The power-law index of the radial cloud distribution has values of $p=-2$, \textbf{0}, $+2$, corresponding to cloud number density distributions $n(r)\propto r^{-4}$, $r^{-2}$, or $r^{0}$, respectively. 
Finally, the angular width of the torus was chosen to represent either a thick disk ($\sigma$ =\textbf{45$^{\circ}$}) or a sphere ($\sigma$ =90$^{\circ}$) since the surface boundary is sharp. 
The bold-faced parameter values are those used  in our ``standard" model. 

The cloud orientation effects described in Section~\ref{sec:direct} are a direct result of radiative transfer in the dust clouds, and are therefore automatically included in most of the models presented here. These should be regarded as the most realistic cases. However, for comparison, and particularly to highlight the effects of cloud orientation, we also present response functions for models that do not include this effect. These cases utilize notional ``isotropically emitting" clouds; the emission from these clouds is computed by averaging the fluxes from the illuminated ($\alpha=180^{\circ}$) and non-illuminated surfaces ($\alpha=0^{\circ}$).

All response functions are the average of five simulation runs, each including 50,000 clouds. The response functions are normalized to their respective maximum values, with the quiescent state luminosity subtracted, so that the response amplitude varies between 0 and 1. Time is expressed in units of the light-crossing time of the torus outer radius, i.e. $2R_{\rm o}/c$.

		\subsection{Spherical Shell Models}\label{sec:spherical}

			\subsubsection{Transfer Function for Blackbody Emission}
In order to isolate the effects on the response function shape of the radial extent ($Y$) of the torus and the radial cloud distribution (determined by the power-law index, $p$), it is useful to consider the transfer function for a spherical shell that contains clouds that radiate isotropically as blackbodies. If the volume emissivity is a power-law function of radial distance from the central continuum source,  $\varepsilon (r)\propto r^{\eta}$, the 1D transfer function is given by \citep{Robinson:1990aa, Perez:1992aa},

\begin{equation}\label{TFana}
\Psi(\tau)  \propto \left\{ \begin{array}{ll}
Y^{\eta+2}-1 &  \tau<1/Y \\
Y^{\eta+2}(1-\tau^{\eta+2}) &    \tau>1/Y,
\end{array}
\right. 
\end{equation}
where $\tau=ct/2R_{\rm o}$. When $\eta=-2$ the transfer function becomes,
\begin{equation}\label{TFana2}
\Psi(\tau)  \propto \left\{ \begin{array}{ll}
\ln(Y) &  \tau<1/Y \\
\ln(\frac{1}{\tau}) &    \tau>1/Y.
\end{array}
\right. 
\end{equation}

The volume emissivity, $\varepsilon$, is the product  of the power radiated by a cloud, $L(r)$, with the cloud number density distribution, $n(r)$. 
For blackbody clouds of constant size,  the bolometric power is $L(r)\propto r^{-2}$ and, as noted in Section~\ref{sec:structure}, $n(r)\propto r^{p-2}$. Thus, in this case, the volume emissivity is $\varepsilon(r)\propto r^{\eta}\propto r^{p-4}$. 

\begin{figure}[t]
\centering
%\begin{center}
	\includegraphics[scale=0.4]{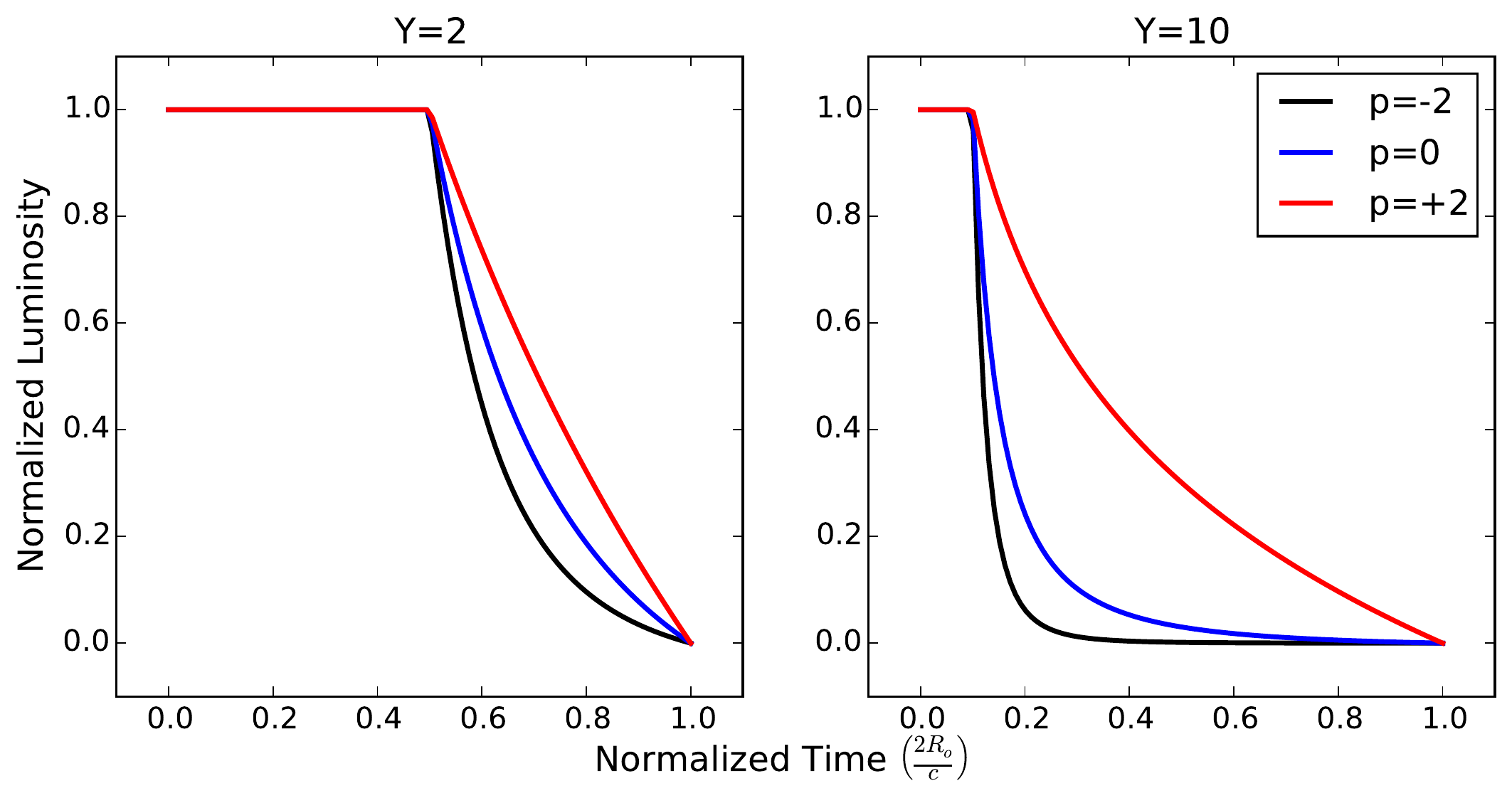}
%\end{center}
\caption{Transfer functions for a spherical shell filled with isotropically emitting blackbody clouds, calculated using equations~\ref{TFana} and~\ref{TFana2}, for $Y =2$ (left) and 10 (right). The different colors show different volume emissivity distributions,  $\varepsilon(r)\propto r^{\eta}\propto r^{p-4}$, for $p = -2, 0, 2$.}
\label{bbsphere}
\end{figure}

The spherical shell transfer function is plotted in Figure~\ref{bbsphere} for $Y = 2$ and $10$, and three values of $p$ $(=-2, 0, 2)$.
As is clear from Equations~\ref{TFana} and~\ref{TFana2}, the transfer function is constant at its maximum amplitude for $0 \leq  \tau \leq 1/Y$, that is,  within the light-crossing time of the inner radius ($2R_{\rm d}/c$). We will henceforth refer to this segment of the response as the ``core".  At times $\tau > 1/Y$, the transfer function decays with a slope that is determined by the value of $\eta$. 
Emissivity distributions (or equivalent in this case, the cloud number density) that decline more steeply with radius produce steeper decays. Henceforth, we refer to the decaying segment of the response as the ``tail''.

As a reference, for comparison with the TORMAC response functions presented below, spherical shell transfer functions corresponding to the same values of $Y$ and $p$ are plotted in 
Figures~\ref{dustysphere2} $-$~\ref{aniso30}.

			\subsubsection{Response Functions for Dust Emission}
\begin{figure}[t]
\begin{center}
	\includegraphics[scale=0.45]{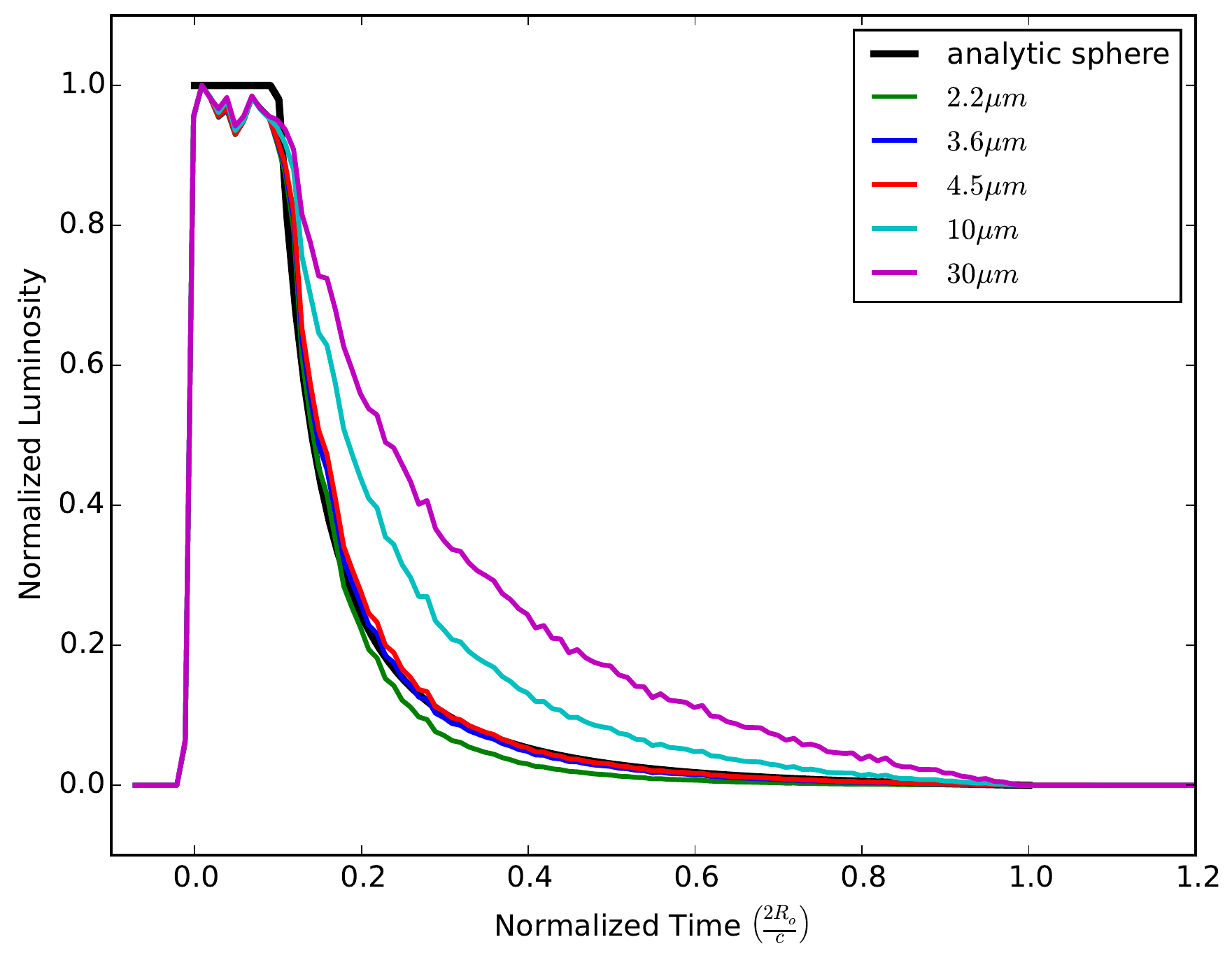}
	\includegraphics[scale=0.45]{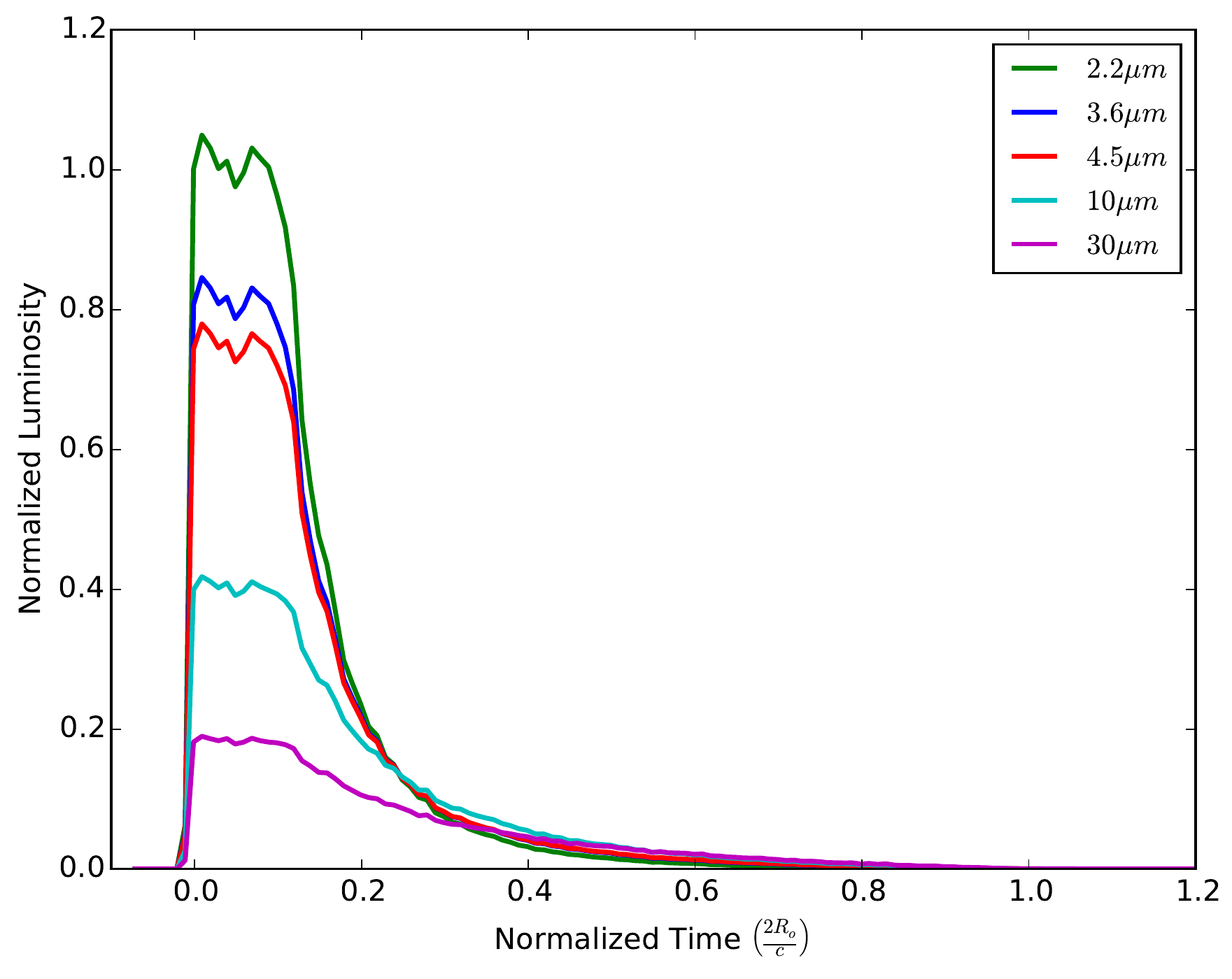}
\end{center}
\caption{Simulated response functions using the pre-computed dust emissivity grid for a spherical dust distribution with isotropic cloud emission, $Y$=10, and $p$=0 at various wavelengths. In the top panel, the simulations are compared to the analytic spherical case for blackbody clouds (solid black line). In the bottom panel, the simulations are normalized to the maximum value at 2.2 $\mu$m in order to show the relative amplitude at different wavelengths.}
\label{dustysphere2}
\end{figure}

It is also of interest to compare the torus responses at different wavelengths, for simulations in which the cloud emission is derived from the radiative transfer model grid, with the corresponding analytical transfer function for blackbody clouds.
The top panel of Figure~\ref{dustysphere2} compares the TORMAC response function at several specific wavelengths (2.2, 3.6, 4.5, 10, and  30 $\mu$m) to the analytic solution for a spherical shell with $Y=10$ and $p=0$.
To facilitate comparison with the analytic transfer function, the cloud orientation effect is not included in the TORMAC models plotted here (i.e., the dust cloud emission is isotropic).

The TORMAC response functions exhibit the same overall shape as the analytical blackbody case. 
However, the steepness of the decay tail varies with wavelength and, in general, no longer follows the blackbody transfer function for the corresponding volume emissivity power-law index derived above ($\eta=p -4$). 
The luminosity of a cloud in this case is not simply proportional to $r^{-2}$, but is dependent on the shape of the dust emissivity curve at that wavelength (Figure~\ref{dusttemp}). For example, at longer wavelengths, the dust emissivity is less sensitive to the cloud surface temperature, which in turn results in a slower decline with radius in the volume emissivity of the cloud ensemble. Thus, the decay of the response function tail becomes shallower as wavelength increases.

The bottom panel of Figure~\ref{dustysphere2} shows the same response functions, but in this case normalized to the maximum value of the 2.2 $\mu$m response function in order to compare the change in luminosity at the different wavelengths. At short delays, $\lesssim 1/Y$, the response amplitudes are much larger at shorter wavelengths, at which the emission is dominated by the hotter, inner clouds. The emission at longer wavelengths from the cooler, outer clouds is more important at longer delays $\ge 1/Y$.
Thus, the hotter dust has more influence on the response at shorter delays, with the cooler dust taking over at longer delays. 

	\subsection{Disk Models}\label{sec:disk}

The simulations for a spherical distribution of dust clouds discussed above serve to illustrate the wavelength dependence of the dust emission response. 
Here, we present response functions for a thick disk geometry ($\sigma = 45^{\circ}$) representing the torus. Although a Gaussian distribution in $\beta$ is probably a more realistic representation of the cloud distribution in altitude above the disk plane, in these models, we adopt a sharp-edged disk geometry for simplicity. Response functions for ``fuzzy'' torus models will be explored in a forthcoming paper.
In fact, our tests show that the distribution in $\beta$ does not have a large effect on the shape of the response functions for the parameter ranges considered here; the response functions for ``fuzzy'' and sharp-edged disks are qualitatively very similar.

\begin{figure}[t]
%\begin{center}
	\includegraphics[scale=0.48]{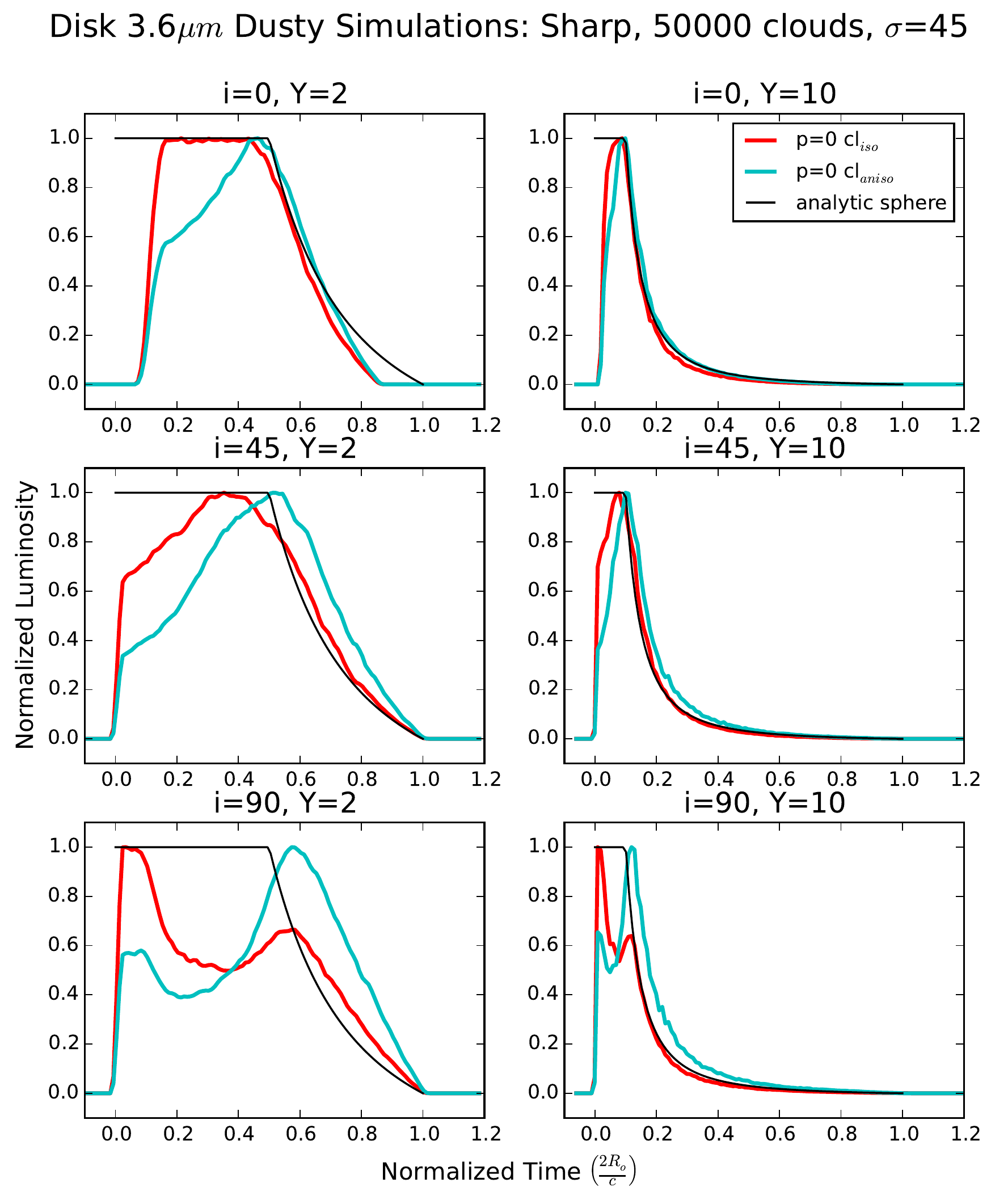}
%\end{center}
\caption{Response functions for an isotropically illuminated torus with $\sigma$=45$^{\circ}$; $Y$=2, 10; $p$=0; and $i=0^{\circ}, 45^{\circ}, 90^{\circ}$ at 3.6 $\mu$m. The blue lines represent the models with cloud orientation and the red lines represent models without cloud orientation for comparison. The black line represents the analytic transfer function for a spherical shell.}
\label{clorient}
\end{figure}

\begin{figure}[t]

	\includegraphics[scale=0.48]{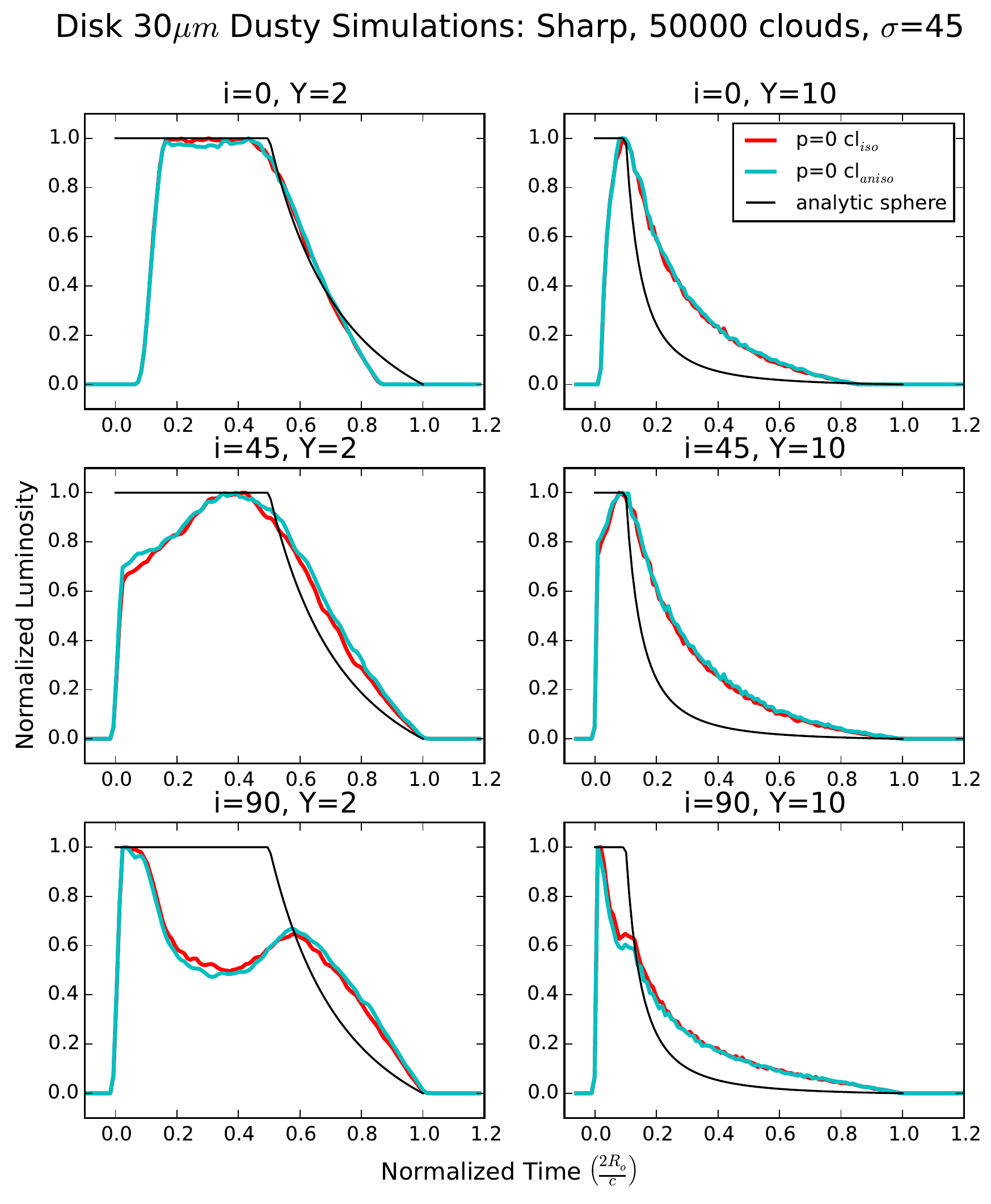}

\caption{Response functions for an isotropically illuminated torus with $\sigma$=45$^{\circ}$; $Y$=2, 10; $p$=0; and $i=0^{\circ}, 45^{\circ}, 90^{\circ}$ at 30 $\mu$m. The blue lines represent the models with cloud orientation and the red lines represent models without cloud orientation for comparison. The black line represents the analytic transfer function for a spherical shell.}
\label{clorient30}
\end{figure}

Models incorporating cloud orientation for cases in which the torus is illuminated isotropically or anisotropically by the central source are compared in Figures~\ref{clorient} $-$~\ref{aniso30}. This set consists of compact ($Y=2$) and radially extended ($Y=10$) torus models, at inclinations $i=0^{\circ}, 45^{\circ},$ and $90^{\circ}$. The power law index of the radial cloud distribution is taken to be $p=0$ in all models. For comparison, we also plot response functions for a corresponding set of models which do not include cloud orientation (red lines) in Figures~\ref{clorient} and~\ref{clorient30}. The analytical spherical shell transfer functions for the same values of $p$ and $Y$ are also plotted for comparison in Figures~\ref{clorient} and~\ref{clorient30}.

		\subsubsection{General Features of the Response}\label{sec:general_features}

Here, we briefly comment on the general features of the disk response function with respect to inclination and radial extent. For this purpose, it is useful to consider the case where the torus is illuminated isotropically and the clouds emit isotropically (i.e., cloud orientation is neglected). The differences that arise as a result of cloud orientation effects and anisotropic illumination will be discussed in Sections~\ref{sec:aniscloud} and~\ref{sec:anisotor}, respectively.

As for the spherical shell, the disk response functions achieve maximum amplitude for delays $\tau \la 1/Y$, and decay at later times $\tau > 1/Y$, with the decay tail usually being shallower at 30 $\mu$m than at 3.6 $ \mu$m 
(this is most evident for $Y=10$). However, the disk response functions also exhibit distinctive features compared to the spherical case. In particular, 
the response function is not constant at its maximum amplitude for delays $0 \leq  \tau \leq 1/Y$, but the form of the response during this period depends on inclination. For a face-on torus ($i=0^{\circ}$), the response functions exhibit a delay before the onset of the response, which results from the fact that there are no clouds within a cone of opening half-angle $90^{\circ}-\sigma,$ aligned with the observer's line of sight. 

At large inclinations, the disk typically exhibits a double-peaked response function. This feature appears because
the isodelay surface intersects fewer clouds as it passes through the cavity at the center of the torus. Therefore, the time delay between the peaks is largely determined by the light crossing time of the inner radius of the torus. For edge-on tori ($i = 90^{\circ}$), the separation between the peaks is $\approx 1/Y$. 
 The first peak has a higher amplitude, and is due to the emission of clouds on the side of the disk nearest to the observer. The second, lower amplitude peak is due to the delayed and smeared emission response from the far side of the disk. 
 The double peaks are not seen for $i$ = 45$^{\circ}$ because the disk is relatively thick in this model ($\sigma = 45^{\circ}$). Thus, at this inclination, there are still many clouds on the isodelay surface as the pulse passes through the torus, so there is a gradual increase in the response rather than a strong initial peak.

			\subsubsection{Effects of Cloud Orientation}\label{sec:aniscloud}

As discussed in Section~\ref{sec:direct}, the flux observed from any cloud is a function of its position angle, $\alpha,$ which in turn is determined by the cloud's location within the torus. This cloud orientation effect has a dramatic influence on the response function at short wavelengths, where the emission from the illuminated side is much stronger than from the non-illuminated side (see Figure~\ref{dusttemp}). 
The clouds on the near side of the torus, which respond after shorter delays, present a larger fraction of their cooler (hence fainter), non-illuminated sides to the observer. 
As a result, the response function is suppressed at short delays for all inclinations, so that the peak amplitude is reached at later times than is the case when cloud orientation is neglected. This effect can be seen in the 3.6 $\mu$m  response functions presented in Figure~\ref{clorient}. At $i = 90^{\circ}$, for example, the initial response function peak has a much lower amplitude than the second peak when cloud orientation is included.

However, although the response function is strongly modified by cloud orientation at 3.6 $\mu$m, the effect is much weaker at longer wavelengths ($>10\, \mu$m), at which the clouds emit almost isotropically (Figure~\ref{dusttemp}).
 As can be seen in Figure~\ref{clorient30}, cloud orientation has little effect on the response functions at 30 $\mu$m.
An interesting consequence of this wavelength dependence is that the response function actually reaches its maximum after a shorter delay at 30 $\mu$m than at 3.6 $\mu$m.

The response function centroids, $\tau_{\rm c}$, are given in Table~\ref{otherresults}; here, we consider only the models in which the AGN radiation field is isotropic (columns 3$-$6). 
In general,  $\tau_{c}$ scales with $1/Y$ because, as already noted, the response functions reach their largest amplitudes for $\tau \la 1/Y$. 
For models without cloud orientation, the response functions have larger centroids at 30 $\mu$m than at 3.6 $\mu$m, particularly for $Y=10$, due to the extended tail. 
There is barely any change with inclination at both wavelengths.

However, when cloud orientation is included in the models, the 
centroids at 3.6 $\mu$m increase for all values of $Y$ and $i$, because the maximum of the response function shifts to larger delays. The centroid also increases slowly with inclination. 
On the other hand, at 30 $\mu$m, the clouds emit almost isotropically; hence, the centroids increase only slightly.  
For $Y=2$, the centroids at 3.6 $\mu$m exceed those at 30 $\mu$m because of the shift in the response function maximum. This does not occur at $Y=10$ because of the extended tail of the 30 $\mu$m response function.
\begin{table}
\caption{{Simulation Centroids}}
\label{otherresults}
\begin{center}
\scriptsize{
\begin{tabular}{lccccccc}
\hline
\hline
$Y$ & $i$ & \multicolumn{6}{c}{$\tau_{\rm c}$ $\left(\frac{2R_{\rm o}}{c}\right)$}\\ 
 \hline
 & &  \multicolumn{2}{c}{cl$_{\rm iso}$; tor$_{\rm iso}$} & \multicolumn{2}{c}{cl$_{\rm aniso}$; tor$_{\rm iso}$} & \multicolumn{2}{c}{cl$_{\rm aniso}$; tor$_{\rm aniso}$} \\
 & & 3.6 $\mu$m & 30 $\mu$m & 3.6 $\mu$m & 30 $\mu$m & 3.6 $\mu$m & 30 $\mu$m \\
\hline
2 & 0$^{\circ}$ & 0.38 & 0.39 & 0.44 & 0.40 & 0.30 & 0.28 \\
 & 45$^{\circ}$ & 0.38 & 0.40 & 0.47 & 0.40 & 0.32 & 0.30 \\
 & 90$^{\circ}$ & 0.38 & 0.39 & 0.53 & 0.41 & 0.34 & 0.30 \\

 \hline
 10 & 0$^{\circ}$ & 0.11 & 0.19 & 0.13 & 0.19 & 0.08 & 0.14 \\
 & 45$^{\circ}$ & 0.11 & 0.18 & 0.14 & 0.19 & 0.08 & 0.14\\
 & 90$^{\circ}$& 0.11 & 0.17 & 0.14 & 0.18 & 0.09 & 0.13 \\
 
\hline
\end{tabular}}
\end{center}
\end{table}

		\subsubsection{Anisotropic Torus Illumination}\label{sec:anisotor}

\begin{figure}[t!]

	\includegraphics[scale=0.48]{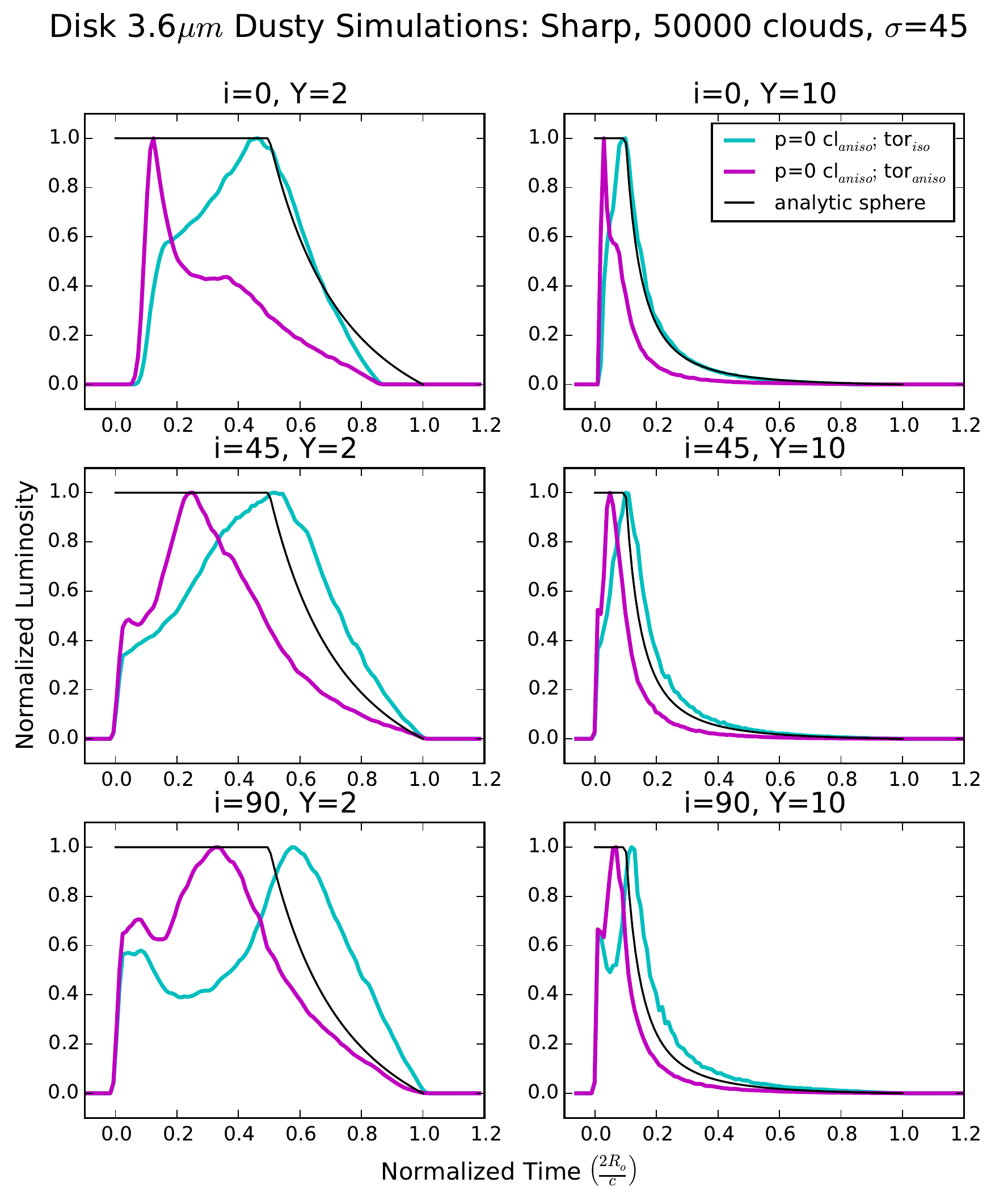}

\caption{Response functions for a torus with $\sigma$=45$^{\circ}$; $Y$=2, 10; $p$=0; and $i=0^{\circ}, 45^{\circ}, 90^{\circ}$ at 3.6 $\mu$m.  The blue and purple lines represent a torus that is illuminated either isotropically or anisotropically, respectively, by the central source. The black line represents the analytic transfer function for a spherical shell.}
\label{aniso}
\end{figure}

\begin{figure}[t!]

	\includegraphics[scale=0.48]{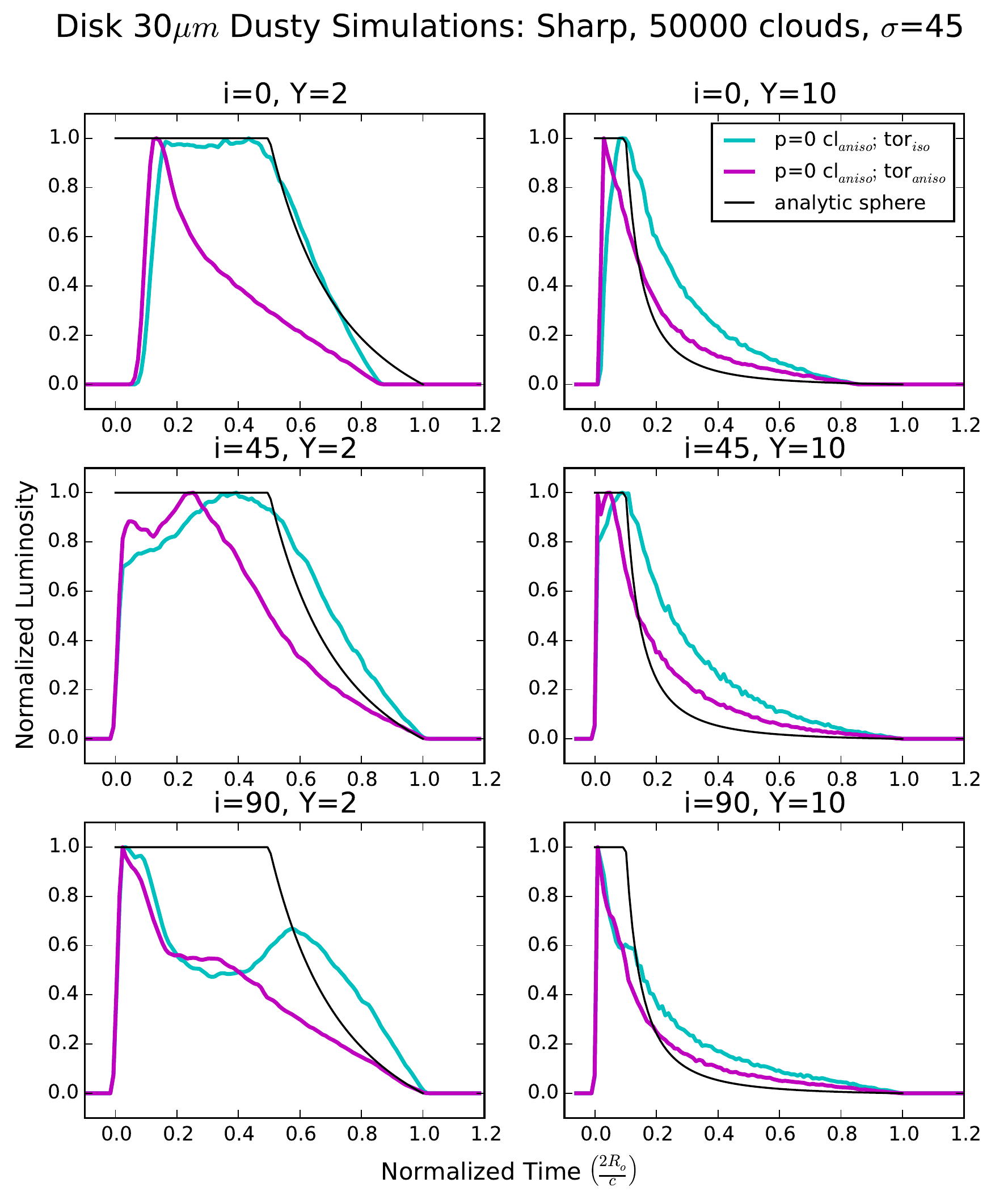}

\caption{Response functions for a torus with $\sigma$=45$^{\circ}$; $Y$=2, 10; $p$=0; and $i=0^{\circ}, 45^{\circ}, 90^{\circ}$ at 30 $\mu$m.  The blue and purple lines represent a torus that is illuminated either isotropically or anisotropically, respectively, by the central source. The black line represents the analytic transfer function for a spherical shell.}
\label{aniso30}
\end{figure}

\begin{figure}[h]
\begin{center}
	\includegraphics[scale=0.5]{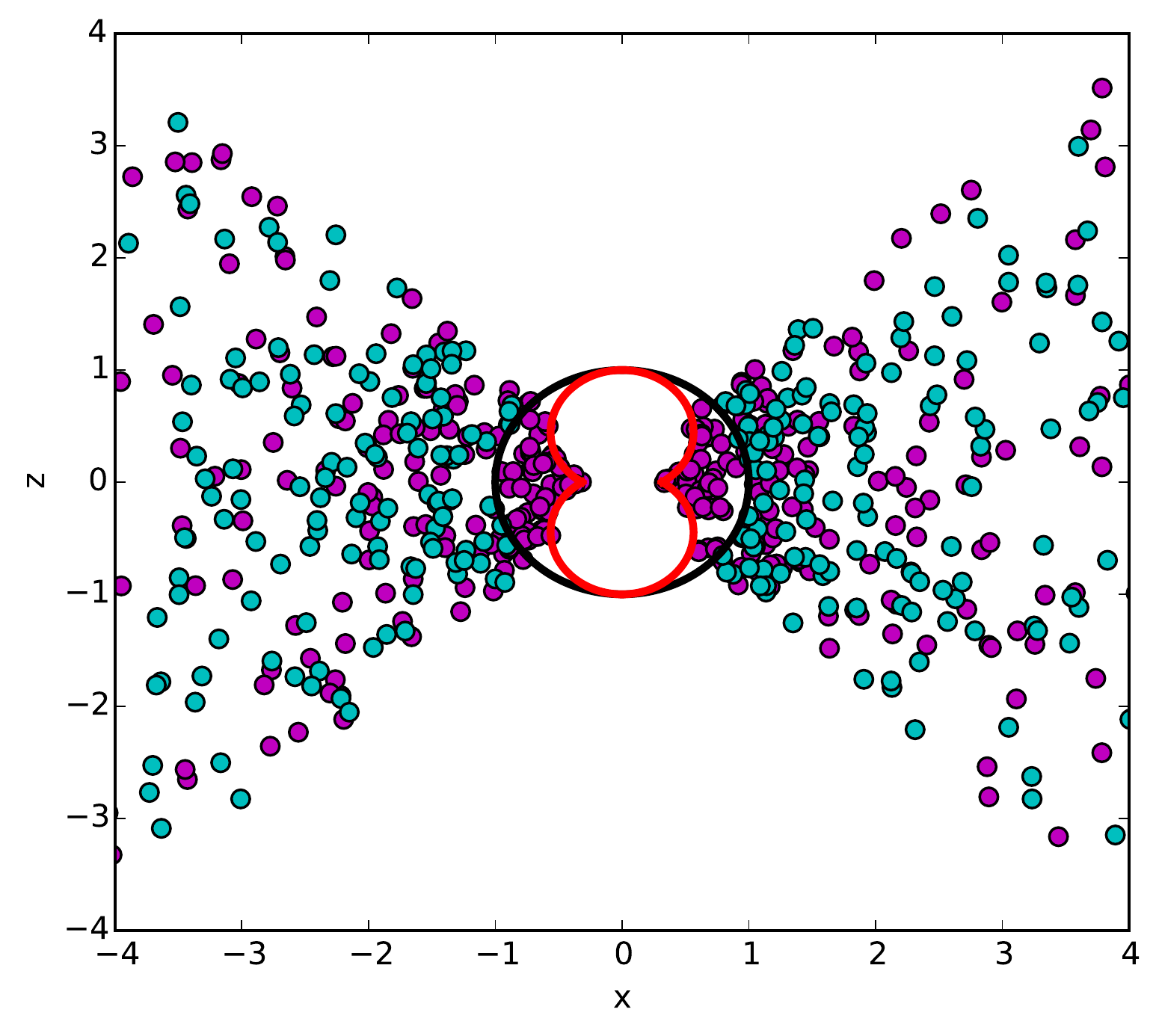}
\end{center}
\caption{A vertical slice through the torus cloud ensemble. The black circle represents the dust sublimation radius for an isotropically emitting point source (blue clouds). These clouds cannot lie within this radius. However, when the AGN radiation field is anisotropic (purple clouds) the dust sublimation radius is a function of polar angle and resembles a figure eight (red line). In this case, clouds can reside closer to the source near the equatorial plane than for the isotropic point source.}
\label{isovsaniso}
\end{figure}

Response functions for models that incorporate both cloud orientation and anisotropic illumination of the torus by the accretion disk are shown in Figures~\ref{aniso} and~\ref{aniso30} for 3.6 $\mu$m 
and 30 $\mu$m, respectively. Also plotted, for comparison, are the response functions for corresponding models with the same geometrical parameters, which include cloud orientation (but with the torus isotropically illuminated). The torus configurations that correspond to the two illumination cases are illustrated in Figure~\ref{isovsaniso}, which shows a vertical slice through the center of the torus.  Anisotropic illumination results in a dust sublimation surface that has a figure eight shape in this slice, which allows dust clouds to reside closer to the AGN continuum source near the equatorial plane.

The response functions of the anisotropically illuminated tori exhibit large differences compared to those of the isotropic illumination models, peaking at shorter lags, and generally exhibiting sharper and narrower features. 
At 3.6 $\mu$m, a double-peaked structure is present for all inclinations, with a shorter delay between the first and second peaks than is the case for isotropic illumination. 
These differences result from the shorter light travel times to the torus inner clouds, which reside at smaller radii near to the equatorial plane. 

It is notable that, for anisotropic illumination in the face-on case, the response reaches its maximum amplitude in a sharp peak after a short delay due to the lack of clouds along the line of sight. In contrast,
the response for isotropic  illumination is suppressed at early times due to the effects of cloud orientation, as discussed in Section~\ref{sec:aniscloud}, and does not reach its maximum until $\tau \sim 1/Y$. 
This occurs because the innermost clouds have larger values of $\alpha$, compared to clouds at larger radii that respond at a similar delay (i.e., more distant clouds located along the same isodelay surface). 
These inner clouds are not only the hottest (hence brightest), but have more of their illuminated surfaces facing the observer than other clouds along the same isodelay surface. The later, secondary peak is due to the response of the clouds furthest from the observer along the anisotropic inner radius of the torus (i.e., clouds with $\theta\ga90^{\circ}$). 

As the cloud emission is essentially isotropic at 30 $\mu$m, the response amplitude at short delays is not suppressed, and the response exhibits an initial peak with a high amplitude at all inclinations. 
However, as for 3.6 $\mu$m, the anisotropically illuminated torus has response functions that peak at shorter lags and have narrower cores than those of the isotropically illuminated torus. 
As can be seen from Table~\ref{otherresults}, the response function centroids for the anisotropically illuminated torus are $30-40$\% smaller than those for the isotropically illuminated torus model with cloud orientation.

\section{Discussion}
In this paper, we are primarily concerned with the effects of cloud orientation and anisotropic illumination of the torus. Accordingly, we have considered only a limited set of geometrical configurations, namely a compact ($Y=2$) and an extended torus ($Y=10$), at face-on, intermediate, and edge-on inclinations ($i=0^{\circ}, 45^{\circ},$ and $90^{\circ}$). However, other structural and geometrical parameters also influence the response; in particular, the angular thickness ($\sigma$) and the radial cloud distribution ($p$). A more comprehensive exploration of the effect that these and other parameters have on the dust emission response function will be presented in a future paper. 

The cloud emission at various wavelengths is determined as a function of cloud surface temperature and cloud orientation angle by interpolation in a grid of radiative transfer models. The radiative transfer calculations introduce two important effects which influence the response function. First, the variation of emitted flux with surface temperature (and hence radius, or time) is strongly wavelength-dependent, with the flux decreasing more rapidly with temperature at shorter wavelengths. Second, the cloud emission is strongly anisotropic at shorter wavelengths, with the difference between the output fluxes of the illuminated and non-illuminated sides of the cloud becoming smaller as wavelength increases.

The first effect can be seen in Figure~\ref{dustysphere2}, which show that the more gradual variation of the emitted flux with cloud surface temperature at longer wavelengths (10 and 30 $\mu$m) results in a slower decay of the response function tail, and a smaller amplitude. 
This is more important for the models with $Y=10$, because at $Y=2$ there is a limited range in cloud surface temperature between the inner ($T=T_{\rm sub}=1500$ K) and outer ($T\approx$1150 K) radii of the torus. For $Y=10$, the surface temperature at the outer radius is $T\approx 620$ K, and the response function exhibits a much more gradual decay for $\tau > 1/Y$ at 30 $\mu$m, than at 3.6 $\mu$m. 

The fact that the cloud emission is anisotropic strongly affects the response functions at shorter wavelengths.
The 3.6 $\mu$m response is suppressed at short delays; in contrast, the 30 $\mu$m response is barely affected. Therefore, the response functions show significant differences, even for $Y=2$. The suppression of the response at short delays also leads to an increase in the response function centroid by up to $\sim40$\%, depending on $Y$ and $i$, which in turn will result in a longer lag at 3.6 $\mu$m than would be the case if the cloud emission were isotropic.

These results suggest that it is important to take into account the anisotropic cloud emission due to cloud orientation, and its wavelength dependence, when
interpreting optical-IR lags derived from time series analysis of light curves at IR wavelengths $\la 10\, \mu$m. Neglecting this effect will lead to an overestimation of the radius of the emitting region. In addition, cloud emission anisotropy also has the effect of making the response function more sensitive to torus inclination at shorter wavelengths, with the centroid increasing as inclination increases.

Also, the optical-IR AGN SED changes with time due to variations of both the driving optical continuum as well as the reverberation response of the IR torus emission. Therefore, even when simultaneous observations are available, one must take into account the reverberation response when
analyzing and modeling SEDs \citep{Fausnaugh:2016aa}. 

Anisotropic illumination of the torus by the AGN continuum also dramatically changes the torus response function. As the dust sublimation radius is, in this case, a function of polar angle, and is smaller near the equatorial plane than at the poles, the response functions exhibit narrower cores, shorter lags, and sharper, more distinct features, regardless of the effects of cloud orientation. Compared to the isotropically illuminated cases, large differences are evident at both 3.6 and 30 $\mu$m. In fact, at 3.6 $\mu$m, anisotropic illumination counteracts the suppression of the response at short delays due to cloud orientation. The response function centroids for the anisotropically illuminated torus are smaller, by $\sim35$\% at 3.6 $\mu$m and $\sim30$\% at 30 $\mu$m, than for the isotropically illuminated torus with cloud orientation. Our results agree with the findings of \cite{Kawaguchi:2010aa,Kawaguchi:2011aa} on the general point that anisotropic illumination of the torus can substantially decrease the lag, and may
at least partly explain the discrepancy between the theoretical sublimation radius and measured K-band and mid-IR reverberation lags \citep{Koshida:2014aa, Vazquez:2015aa}.

\subsection{Light Curve Simulations: NGC 6418}
	
\begin{figure}[t]
%\begin{center}
	\includegraphics[scale=0.47]{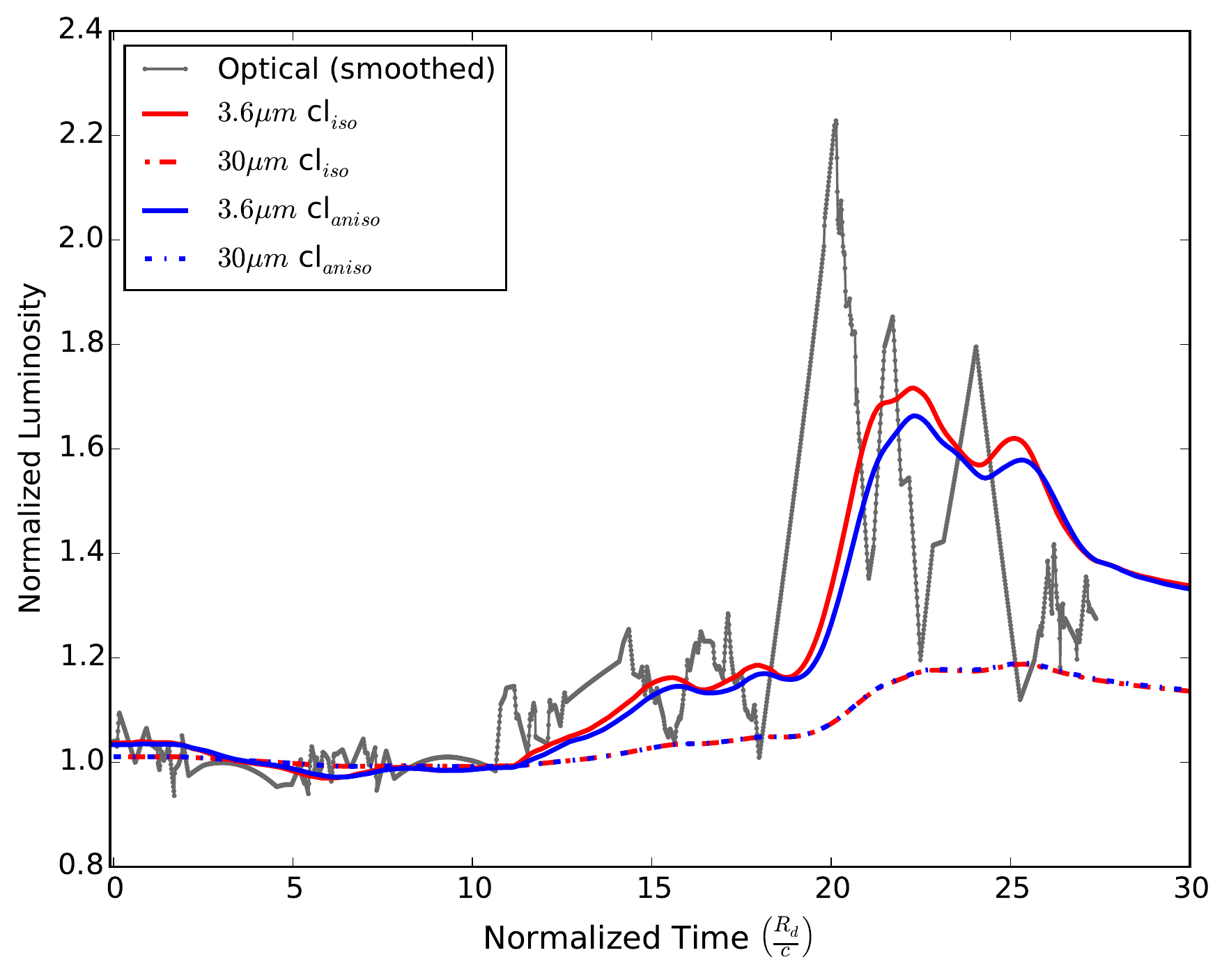}
%\end{center}	
\caption{Simulations of IR light curves (solid=3.6 $\mu$m and dashed-dotted=30 $\mu$m lines) for the observed optical light curve of NGC 6418, interpolated using a damped random-walk model (gray points and gray solid line). These simulations are for an isotropically illuminated torus without cloud orientation taken into account in red, and with cloud orientation in blue. The torus is made up of 50,000 clouds and is viewed face-on with $Y=10$, $p=0$, and $\sigma$=45$^\circ$.}
\label{NGC6418}
\end{figure}

So far, only response functions (the responses to a single, short optical pulse) have been examined in detail. As discussed in Section~\ref{sec:resfuncs}, the response functions display different signatures, which vary with wavelength, due to the cloud orientation effect, and also differ between isotropically and anisotropically illuminated tori.
However, can these signatures be identified in observed light curves, and what effect will they have on the lags recovered from time series analysis?

\begin{figure}[t]

	\includegraphics[scale=0.46]{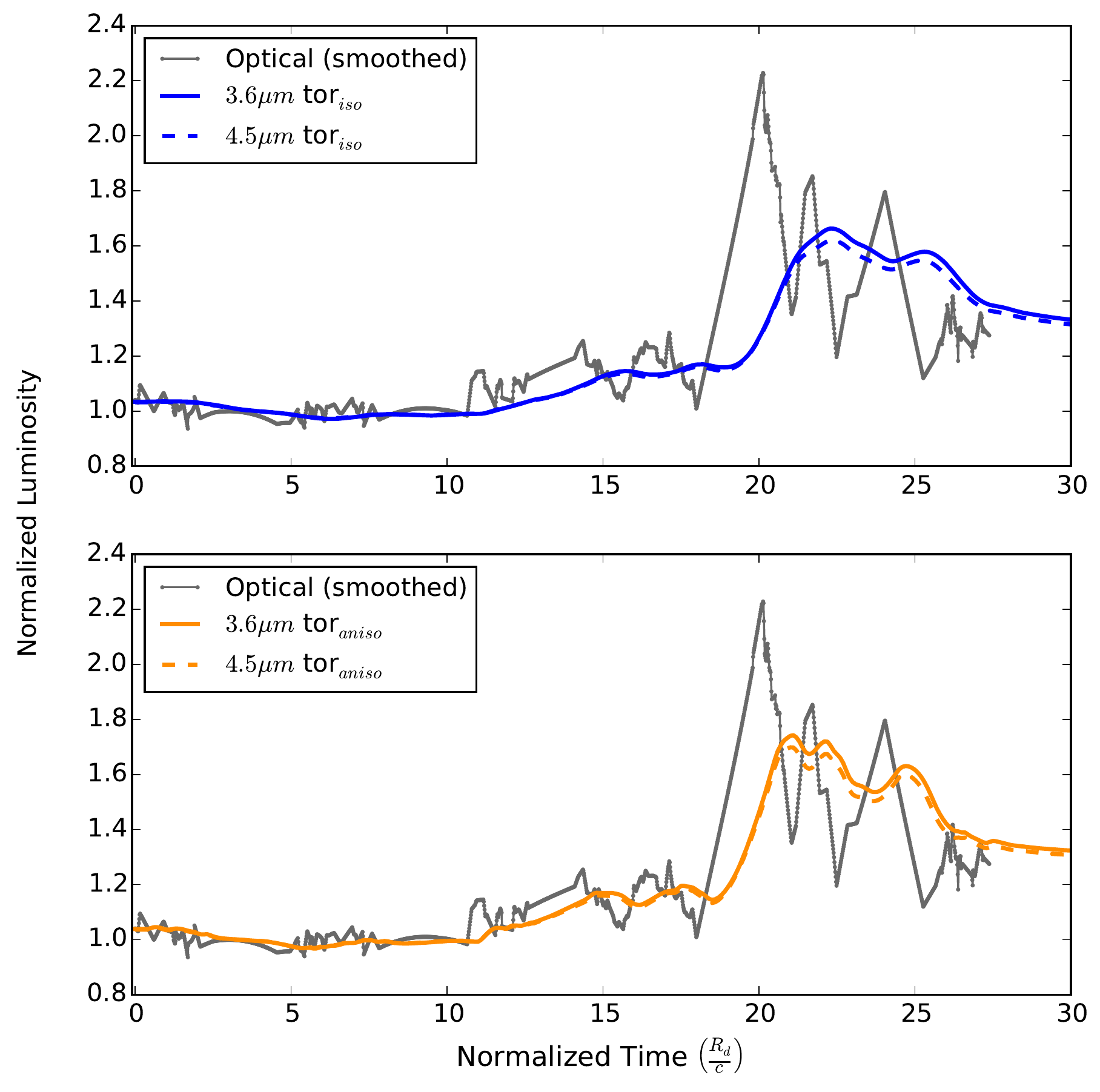}

\caption{Simulations of IR light curves (solid=3.6 $\mu$m and dashed=4.5 $\mu$m lines) for the observed optical light curve of NGC 6418, interpolated using the damped random-walk model (gray points and gray solid line). The isotropically illuminated torus is shown in blue in the top panel. The anisotropic case is in orange, in the bottom panel. The torus is made up of 50,000 clouds and is viewed face-on with $Y=10$, $p=0$, and $\sigma$=45$^\circ$.}
\label{NGC6418twopan}
\end{figure}

To gain some insight into these questions, we have used TORMAC to compute simulated IR light curves for selected models using the optical light curve of NGC 6418, one of the AGN observed during the Spitzer monitoring campaign, as the input. This light curve was compiled from new observations supplemented with data obtained from the Catalina Sky Survey \citep{Drake:2009aa} and the Palomar Transit Factory \citep{Law:2009aa}. The observations and measurements are described in \cite{Vazquez:2015aa}, \cite{Vazquez:2015ab}, and A. Robinson et al., in preparation. In order to interpolate over gaps in the data, and hence generate a regularly sampled light curve for use as the input for TORMAC, the observed optical light curve was reprocessed using the damped random walk model (JAVELIN) from \cite{Zu:2011aa}. The input optical light curve (shown in Figures~\ref{NGC6418} and~\ref{NGC6418twopan}) is a weighted mean of many realizations consistent with the observed data. It is characterized by posterior density distributions for the damping timescale and the amplitude, which have medians of 175 days and 0.24 in relative flux units, respectively.

The input optical light curve and simulated IR light curves are shown in Figures~\ref{NGC6418} and~\ref{NGC6418twopan}. It should be noted that our intent in presenting these simulations is not to fit the observed IR light curves, but simply to illustrate the effects of cloud orientation and anisotropic illumination of the torus on the IR response light curves.

In these simulations, the torus is composed of 50,000 clouds and is viewed face-on ($i=0^\circ$) with $Y=10$, $p=0$, and $\sigma$=45$^\circ$. Cloud shadowing is not taken into account. In Figures~\ref{NGC6418} and~\ref{NGC6418twopan}, time is given in units of $R_{\rm d}/c$ =70.8 days (i.e., half the light-crossing time of the inner radius of the torus). The value of $R_d$ was determined from Equation~\ref{dustsub}, using the bolometric luminosity of NGC 6418 as determined by \cite{Vazquez:2015aa}, $L_{\rm AGN}=2.21\times10^{43}$ erg s$^{-1}$. The luminosity of each light curve is normalized to that of its initial state. 

Figure~\ref{NGC6418} shows the simulated 3.6 and 30 $\mu$m light curves for a torus that is illuminated isotropically (blue lines). For comparison, the corresponding model in which cloud orientation is neglected (i.e., all clouds are forced to emit isotropically; red lines) is also shown.
The 3.6 $\mu$m light curves exhibit similar variations in response to the optical continuum, but also show clear differences in that the light curve for the cloud orientation case has a lower amplitude and is more smeared out. This can be attributed to the fact that, on average, the clouds on the near side of the torus have a larger fraction of their cooler, non-illuminated side facing the observer.

As would be expected from the response functions, the 30 $\mu$m light curves are essentially identical for the two cases, a consequence of the fact the cloud emission is almost isotropic at this wavelength. The 30 $\mu$m light curves also show a much smaller changes in luminosity, and are much more smeared out than the optical or the corresponding 3.6 $\mu$m light curves. As discussed in Section~\ref{sec:resfuncs}, the slower response and smaller variation amplitude result from the relatively shallow gradient of the emissivity -- temperature curve at this wavelength (Figures~\ref{dusttemp} and~\ref{dustysphere2}).

\begin{table}
\caption{{Simulation Lags: NGC 6418}}
\label{lags}
\begin{center}
\scriptsize{
\begin{tabular}{lccc}
\hline
\hline
 $\lambda$ ($\mu$m)& \multicolumn{3}{c}{$\tau_{\rm lag}$ (days)}\\ 
 \hline
 & cl$_{\rm iso}$; tor$_{\rm iso}$ & cl$_{\rm aniso}$; tor$_{\rm iso}$ & cl$_{\rm aniso}$; tor$_{\rm aniso}$ \\
\hline
3.6 $\mu$m & 125.80$^{+1.03}_{-1.45}$ & 146.28$^{+1.47}_{-1.48}$ & 84.18$^{+1.01}_{-1.47}$ \\
4.5 $\mu$m & 128.81$^{+1.48}_{-1.47}$ & 143.79$^{+1.03}_{-1.48}$ & 79.67$^{+1.01}_{-1.45}$ \\
30 $\mu$m &  215.02$^{+2.99}_{-2.51}$ & 216.01$^{+3.38}_{-3.01}$ & 160.92$^{+2.51}_{-2.45}$ \\
 
\hline
\end{tabular}}
\end{center}
\end{table}

In Table~\ref{lags}, we list optical--IR lags determined by cross-correlating the simulated IR light curves with the input optical light curve, using the method described by \cite{Vazquez:2015aa}. 
At shorter wavelengths, the models that include cloud orientation have significantly longer lags than those that do not. 
Furthermore, the lag at 3.6 $\mu$m in this case is actually slightly longer than that at 4.5 $\mu$m. Again, this occurs because the near-side clouds, which respond first, have more of their cooler non-illuminated sides facing the observer, and are therefore brighter at 4.5 $\mu$m than at 3.6 $\mu$m. As would be expected from the response functions, the lag at 30 $\mu$m is essentially unchanged by cloud orientation.

Figure~\ref{NGC6418twopan} compares the simulated IR light curves at 3.6 and 4.5 $\mu$m for models in which the torus is either illuminated isotropically (top panel) or anisotropically (bottom panel) by the central source. In both cases, cloud orientation (anisotropic dust cloud emission) is included. As seen in the corresponding response functions, the anisotropic illumination model yields shorter response times, larger amplitudes, and is less smeared out. The 3.6 and 4.5 $\mu$m light curves are very similar in shape, as expected from their dust emissivity curves (see Figures~\ref{dusttemp} \&~\ref{dustysphere2}). 
This behavior is reflected in the lags, which are $\sim$40\% shorter at 3.6 and 4.5 $\mu$m and $\sim$30\% shorter at 30 $\mu$m than in the isotropic illumination case. The difference is smaller for the isotropically illuminated torus without cloud orientation. %The difference is smaller for the model without cloud orientation and the torus is isotropically illuminated.  

Note that all of the lags for the simulated light curves are much longer than the measured lags for NGC 6418, as reported in \cite{Vazquez:2015aa}. The light curves analyzed by \cite{Vazquez:2015aa} were obtained during Spitzer Cycle 8, and thus correspond only to a portion of the optical light curve presented in Figures~\ref{NGC6418} and~\ref{NGC6418twopan} (roughly that between 11 and $18R_{\rm d}/c$ in our time units). Nevertheless, the measured lags for that period were $\approx$37 and 47 days at 3.6 and 4.5 $\mu$m, respectively. Clearly, the limited set of models presented here do not reproduce these short lags; as already noted, that was not the intention. The models are scaled to the sublimation radius calculated for the standard ISM dust composition and, like the results of K-band reverberation mapping studies (e.g., \citealp{Kishimoto:2007aa}), the reverberation radii measured for NGC 6418 are a factor of $\sim$2 smaller than the theoretical value predicted by Equation~\ref{dustsub}. We defer detailed modeling of the NGC 6418 light curves to a later paper. 

However, it is worth pointing out here that the effect of cloud orientation is important at these wavelengths, and will result in longer lags that will tend to increase the discrepancy between the measured reverberation radii and $R_{\rm d,ISM}$. On the other hand, anisotropic illumination of the torus would result in shorter lags, and thus act to decrease the discrepancy. In addition, the cloud orientation case produces a {\em shorter} lag at 4.5 than at 3.6 $\mu$m, the opposite of what is observed in NGC 6418.

	\subsection{Effects of Cloud Shadowing}\label{sec:shadowing}

\begin{figure}[t!]
\begin{center}
	\includegraphics[scale=0.7]{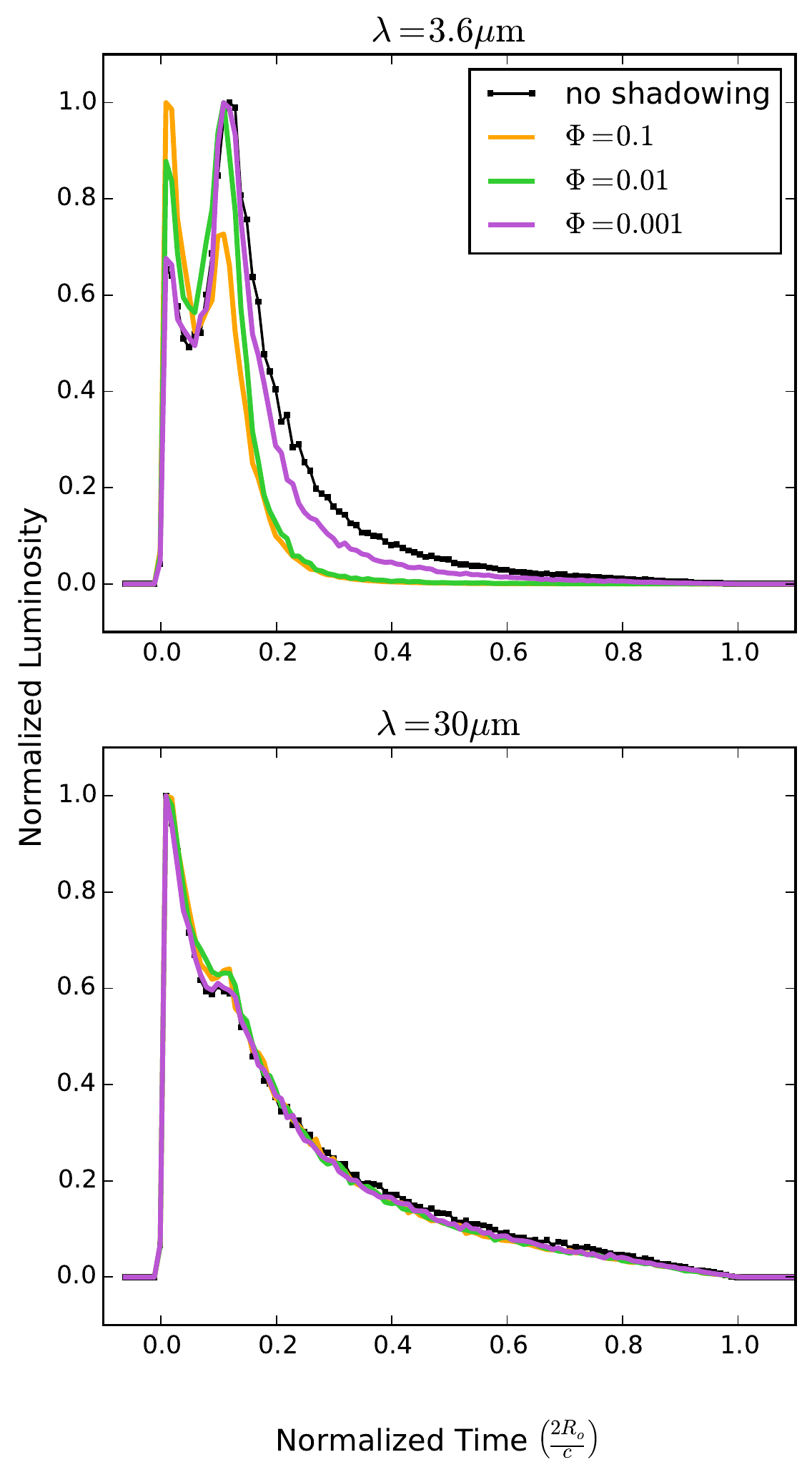}
\end{center}
\caption{Effects of cloud shadowing. Response functions for an isotropically illuminated torus with $\sigma$=45$^{\circ}$; $Y$=10; $p$=0; and $i$=90$^{\circ}$ at 3.6 and 30 $\mu$m.  The black squares and line represent simulations without cloud shadowing and the solid lines represent the simulations with cloud shadowing for values of the average volume filling factor ranging from: $\Phi$=0.001 (purple) to 0.1 (light blue).} 
\label{clshadow_vff}
\end{figure}

\begin{figure}[t]
\begin{center}
	\includegraphics[scale=0.7]{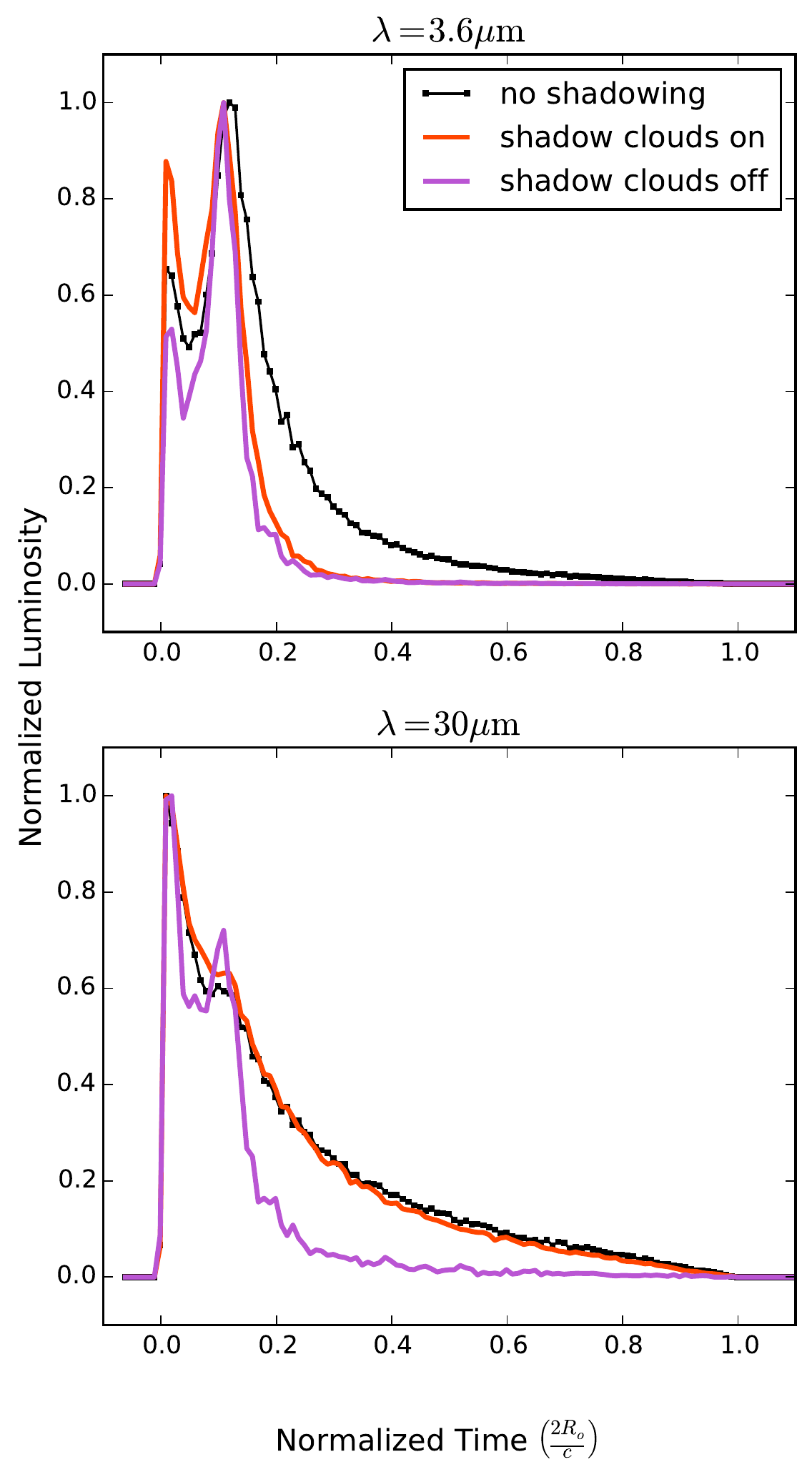}
\end{center}
\caption{Contributions of directly and indirectly heated clouds when cloud shadowing is included. The response functions shown are for an isotropically illuminated torus with $\sigma$=45$^{\circ}$; $Y$=10; $p$=0; and $i$=90$^{\circ}$ at 3.6 and 30 $\mu$m.  The black squares and line represent simulations without cloud shadowing. The solid lines represent simulations where the shadowed clouds were included as indirectly heated clouds (red), or simply excluded (pink). The average volume filling  factor in these cases is $\Phi = 0.01$.} 
\label{clshadow_no_ind}
\end{figure}

As discussed in Section~\ref{sec:clshadow}, there is a high probability that clouds within the torus will be shadowed (i.e., shielded from the AGN continuum source) by clouds nearer to the inner edge of the torus. The shadowed clouds are heated by the diffuse radiation field of nearby directly heated clouds. This effect was not included in the simulations presented in Section~\ref{sec:resfuncs}, because we chose to focus this paper on the effects on the torus response due to cloud orientation and anisotropic illumination of the torus by the AGN continuum. However, it is important to note that cloud shadowing can also significantly modify the torus response.

The fraction of clouds that are shadowed is a function of $Y$, $p,$ and the average volume filling factor $\Phi$, because these determine, respectively, the path length, the degree to which the clouds are centrally concentrated, and the cloud size. As an example, Figure~\ref{clshadow_vff} shows the response functions when cloud shadowing is introduced  for one of the models presented in Section~\ref{sec:resfuncs} (isotopic illumination, cloud orientation, $i=90^{\circ}$, $Y$=10, and $p$=0), for several values of $\Phi$. The average number of clouds along an equatorial ray in the cases shown is $\mathcal{N}(r=R_{\rm o},\beta=0)\approx$22, 5, and 1, for $\Phi$=0.1, 0.01, and 0.001, respectively. In Figure~\ref{clshadow_no_ind}, we compare response functions for $\Phi$=0.01 when both directly and indirectly heated clouds are included, as well as when only directly heated clouds are included (i.e., the shadowed clouds are ``turned off''). In both figures, the corresponding case that does not include cloud shadowing is also plotted for comparison.

As seen in Figure~\ref{clshadow_vff}, cloud shadowing has a dramatic effect at 3.6 $\mu$m, but barely changes the response function at 30 $\mu$m. Without cloud shadowing, the initial peak in the response at 3.6 $\mu$m is suppressed due to the anisotropic emission of the clouds (Section~\ref{sec:aniscloud}). However, with cloud shadowing, this peak is stronger; for $\Phi = 0.1$, it even exceeds the second peak in amplitude. This is because, for higher values of $\Phi$, nearly all of the clouds in the torus interior are shadowed (99.5\% for $\Phi=0.1$). Only clouds near the inner edge are directly heated and emit anisotropically, so the response is dominated by isotropically emitting indirectly heated clouds. This can be seen in Figure~\ref{clshadow_no_ind}, where the initial peak has a much lower amplitude when only the directly heated clouds are included. In addition to changing the relative amplitudes of the peaks in the core for $\tau \leq 1/Y$, cloud shadowing also causes the response function to decay more steeply for $\tau > 1/Y$. Again, this is due to the lack of directly heated clouds at larger radii within the torus. 

As the average volume filling factor is decreased, the response functions change slowly back to the case where cloud shadowing was not included, i.e., the decay tail becomes shallower and the secondary peak increases in height (Figure~\ref{clshadow_vff}). The case without cloud shadowing is recovered for $\Phi \lesssim 0.0001$.

At 30 $\mu$m, as already noted, the introduction of cloud shadowing has a relatively small effect on the response function. This is because at this wavelength, both the directly and indirectly heated clouds essentially radiate isotropically. When the indirectly heated clouds are ``turned off'' (Figure~\ref{clshadow_no_ind}), the emission mostly comes only from the inner edge of the torus for larger values of $\Phi$ and therefore double peaks appear in the core, while the decay tail steepens. 

The response function centroids listed in Table~\ref{vffcompare} reflect these results. At 3.6 $\mu$m, the centroid for $\Phi = 0.1$ is 36\% smaller than for the case where cloud shadowing is not included, due to the steeper decay. However, as $\Phi$ is decreased, the centroid increases, recovering the no cloud shadowing value by $\Phi=0.0001$. On the other hand, at  30 $\mu$m, there is only a  $\approx 10$\% difference in the centroids between the $\Phi = 0.1$ and no cloud shadowing cases.   

It should be noted that the contribution of the indirectly heated clouds is likely overestimated at higher values of $\Phi$, since in the approximation adopted here, the diffuse radiation field is assumed to come from neighboring directly heated clouds (Section~\ref{sec:indirect}). In fact, due to shadowing at higher values of $\Phi$, the majority of directly heated clouds will be found close to the inner edge of the torus. Therefore, shadowed clouds in the interior of the torus are heated by a diffuse radiation field produced predominantly by the inner, directly heated clouds. Thus the shadowed clouds will be heated by a diffuse radiation field which is attenuated by geometrical dilution and extinction. There will also be significant light travel delays between the inner, directly illuminated clouds and the shadowed clouds at larger radii.

\begin{table}
\caption{Simulation Centroids: Cloud Shadowing}
\label{vffcompare}
\begin{center}
\scriptsize{
\begin{tabular}{lccccccc}
\hline
\hline
& $\Phi$  & \multicolumn{2}{c}{$\tau_{\rm c}$ $\left(\frac{2R_{\rm o}}{c}\right)$}\\ 
 \hline
 & &  \multicolumn{2}{c}{cl$_{\rm aniso}$; tor$_{\rm iso}$} \\
 & &  3.6 $\mu$m & 30 $\mu$m  \\
\hline
Without Cloud & 0.1 &  0.14 & 0.18  \\
Shadowing & & &\\
\hline
 With Cloud & 0.1  & 0.09 & 0.16 \\
%& & & \\
 Shadowing & 0.01 & 0.10 & 0.16   \\
 & 0.001 & 0.13 & 0.17  \\
 & 0.0001 & 0.14 & 0.17 \\

\hline
\end{tabular}}
\end{center}
\end{table}

	\subsection{Comparison to other Reverberation Mapping Torus Models} 

TORMAC is very similar in concept to the pioneering torus reverberation model developed by \cite{Barvainis:1992aa}, in that the torus is modeled as an ensemble of optically thick clouds and that the dust emission is determined from a pre-computed grid of radiative transfer calculations. However, TORMAC employs a more sophisticated radiative transfer model (namely, the DUSTY model grid used in the CLUMPY torus model of \citealp{Nenkova:2008aa,Nenkova:2008ab}) and includes cloud orientation and anisotropic illumination by the AGN continuum.

 \citet{Barvainis:1992aa} is mainly concerned with short wavelengths (his models were computed for the J, H, K, and L bands). Nevertheless, the response functions produced by TORMAC for comparable torus models, that is, for an isotropically illuminated torus without cloud orientation, exhibit generally similar features, notably, the delayed onset of the response for a face-on disk and the double-peak response for inclined disks. These are characteristic features of the reverberation response of an annular disk (see also \citealp{Perez:1992aa}). 

More recently, \cite{Kawaguchi:2010aa,Kawaguchi:2011aa} discussed a torus reverberation model that was constructed to study the effects on the NIR response of anisotropic illumination, cloud orientation, and also ``torus self-occultation" (i.e., absorption of NIR emission from dust clouds by other dust clouds). However, the \cite{Kawaguchi:2011aa} model evidently does not explicitly utilize radiative transfer calculations, but instead uses approximate prescriptions for cloud orientation effects and self-occultation. To model the latter effect, in particular, they argue that the torus is effectively optically thick to its own NIR radiation. Hence, they only consider the emission from the near-side surface layer of the torus. 

On the other hand, TORMAC does not currently include self-occultation (although it does account for occultation of the AGN continuum by intervening clouds), and thus the dust emission produced by all clouds is assumed to escape the torus. Therefore, we should not expect detailed agreement in the resulting response functions. Nevertheless, for comparable models, the TORMAC response functions show similar features attributable to cloud orientation (i.e., lower response amplitudes  at shorter delays) and anisotropic illumination (shorter lags/centroids) to those appearing in the \cite{Kawaguchi:2010aa,Kawaguchi:2011aa} models. The main discrepancy appears in the face-on anisotropically illuminated cases (Figure~\ref{aniso}), for which the \cite{Kawaguchi:2011aa} transfer function is sharply single-peaked and narrow, whereas the corresponding TORMAC response functions show an initial sharp peak, followed by a lower amplitude shoulder. This is likely due to the fact that, in the \cite{Kawaguchi:2011aa} model, the response from the far side of the torus is neglected because it is considered to be optically thick to its own NIR radiation.

	\subsection{Current Limitations of TORMAC} 
	TORMAC is capable of computing the multiwavelength response of the AGN torus for parameters governing the geometrical configuration and cloud distribution, and a given dust composition. It includes the effects of cloud orientation and anisotropic illumination. It also accounts for cloud shadowing and includes an approximate treatment of the heating of the shadowed clouds by diffuse IR emission. However, a number of further refinements are necessary to fully capture the complexities associated with self-occultation, dust sublimation, and distributions in cloud optical depth and dust composition.

Arguably the most important limitation of the version of TORMAC used in this paper is that it does not account for self-occultation effects. That is, the IR radiation produced by every cloud is currently assumed to escape the torus without being attenuated by absorption due to any intervening clouds along the line of sight. This is clearly an oversimplification at shorter IR wavelengths, especially in the K-band ($2.2\, \mu$m), where the optical depth is $\tau_{\rm K}\approx4$ for the clouds used in the models presented here (for which, $\tau_{\rm V} = 40$). We anticipate that, for higher values of $\Phi$, thicker tori (i.e., large $\sigma$), and larger inclinations, self-occultation will significantly modify the response function at shorter wavelengths, as found by \cite{Kawaguchi:2011aa}. This would also occur at $10\, \mu$m, where the optical depth is similar to that at 2.2 $\mu$m, due to the Silicate feature. 
Except for very high $\Phi$ values (the case considered, in effect, by \cite{Kawaguchi:2011aa}), the main effect will be to suppress the response at longer delays, because clouds on the side of the torus farthest from the observer are more likely to suffer attenuation. The next version of TORMAC will incorporate wavelength-dependent self-occultation on a cloud-by-cloud basis. The effects on the response function at various wavelengths will be explored in detail in our second paper.

The radiative transfer grid that is used to compute the emission from indirectly heated (i.e., shadowed clouds) is based on a ``local'' approximation for the diffuse radiation field, i.e., it is assumed to come from neighboring directly heated clouds. As noted in Section~\ref{sec:shadowing}, this will tend to overestimate the contribution to the total emission from indirectly heated clouds at higher values of the average volume filling factor, $\Phi$, because most directly heated clouds will be located close to the inner edge of the torus, and there will be relatively few directly heated neighbors within the torus at larger radii. 
In addition, the contribution of the diffuse radiation field to heating directly illuminated clouds is neglected. Heating of their non-illuminated sides by diffuse radiation will tend to reduce the effects of cloud orientation (i.e., reducing how anisotropically the cloud emits) by preferentially increasing the emitted flux from the non-illuminated side.

	TORMAC currently treats every cloud as having the same size, dust composition, and optical depth. However, there are physically motivated reasons to expect a distribution in cloud size (e.g., \citealp{Krolik:1988aa, Beckert:2004aa, Elitzur:2006ab}). For the reasons outlined below, we may also expect gradients in grain composition and size within the torus. It would be desirable, therefore, to implement the capability to handle distributions in these properties.

	It will also be necessary to implement a more sophisticated treatment of time-dependent dust sublimation. The inner radius of the torus is currently defined as the dust sublimation radius. However, as the AGN varies in luminosity, the dust sublimation radius will move in or out, causing the inner radius of the torus to vary depending on the duration and amplitude of the AGN continuum variations, as well as the timescales over which dust clouds are destroyed by sublimation. 
Dust clouds may not be completely destroyed during bright flares in the continuum, but will be eroded by sublimation of grains near the illuminated surface, leaving smaller clouds with lower optical depths. For instance, in the case of NGC 6418 discussed above, the AGN luminosity increased by about 100\% during a flare of total duration $\sim$280 days. A cloud at the sublimation radius before the flare has an illuminated surface temperature of 1500 K. The 100\% luminosity increase would cause the illuminated surface of the cloud to increase to a temperature of about 1700 K. However, the cloud only reaches this temperature for a relatively short period at the peak of the flare, so we assume that its average temperature is 1600 K (corresponding to half the flare maximum). At this temperature, a grain of size 1 $\mu$m sublimates in $\sim$57 days \citep{Waxman:2000aa}. Because a layer of  $\tau_{\rm V} \sim1$ will be eroded from the cloud surface in this time, clouds of  $\tau_{\rm V} \ga 5$ will survive. 
	
	In the current treatment, clouds are not destroyed, but the surface temperature is held constant at the sublimation temperature of 1500 K. The emission from clouds within the current sublimation radius therefore saturates, whereas in fact, grains can survive at higher temperatures for a limited time, leading to higher variability amplitudes at shorter wavelengths. In addition, the sublimation temperature, and therefore rate, is sensitive to dust grain size and composition. Larger grains and those composed of more refractory materials (such as graphite) will survive longer. Therefore, inner dust clouds that survive several continuum variability events are likely to evolve a grain composition that is dominated by larger, more refractory grains than those in the body of the torus \citep{Perna:2003aa}. This would lead to a population of hotter clouds, composed of large, carbonaceous grains, residing within the time-averaged sublimation radius ($r < \langle R_{\rm d} \rangle$), whose emission spectrum is brighter at shorter (NIR) wavelengths.

\section{Conclusions}

This paper has introduced a new dust reverberation mapping model that simulates the response of the IR dust emission from a clumpy AGN torus to variations in the UV/optical luminosity from the central source. 
We present examples of the torus response function (i.e., the IR response to a square wave pulse, an approximation to the transfer function) for ranges in certain geometrical and structural parameters, such as inclination and radial extent, in order to explore how these parameters affect the dust emission response at selected wavelengths. There is a large parameter space that we can cover; however, we chose to focus this paper on investigating the effects of dust cloud orientation and anisotropic illumination of the torus by the AGN accretion disk. 
We also briefly discuss the effects of cloud shadowing, that is, the occultation of the AGN continuum source by intervening clouds.

The advantage of TORMAC is that the combination of a three-dimensional cloud ensemble with a pre-computed grid of radiative transfer emissivities provides the flexibility to explore the multiwavelength response functions for a wide range of torus parameters and dust properties, at a relatively modest cost in computer time. Previous models either do not incorporate cloud orientation or anisotropic illumination, or they employ parametrized dust emissivities, which do not account for the detailed dependence on wavelength.

The code uses a grid of pre-computed radiative transfer models to determine the flux emitted by individual clouds as functions of surface temperature (itself a function of time and position within the torus) and orientation with respect to the observer. This affects the IR response in two important ways. First, the cloud's flux as a function of surface temperature is highly wavelength-dependent, with the emission at shorter wavelengths decreasing more rapidly as temperature decreases. Second, dust clouds emit more anisotropically at shorter wavelengths. As a result, the observed emission from a cloud depends on its orientation with respect to the central source and the observer. Including this cloud orientation effect dramatically alters the response function and causes an increase in the response function centroid (implying an increased lag) for wavelengths $<10\, \mu$m. These effects have important consequences for the IR response, so they must be taken into account when interpreting measured lags from reverberation mapping studies. Indeed, they should also be considered in SED modeling, because the dust emission lags are wavelength-dependent and IR SEDs are usually assembled from non-simultaneous observations. 

Anisotropic illumination of the torus by the AGN continuum source should naturally occur as a consequence of edge darkening, if the continuum is emitted by a geometrically thin accretion disk. This also has a significant effect on the torus response function, dramatically decreasing the IR response lags in all cases. The decrease in lag due to anisotropic illumination overpowers the increase that arises from cloud orientation. As previously discussed by \cite{Kawaguchi:2010aa,Kawaguchi:2011aa}, this illumination effect can partly resolve the discrepancy found when comparing lags measured by reverberation mapping studies to the predicted inner torus radius, if the latter is determined by sublimation and the dust has the standard ISM composition.

Also, we briefly consider cloud shadowing effects, where the shadowed clouds are heated indirectly by an isotropic diffuse radiation field determined by averaging the emission from directly heated clouds at the same distance from the central source. In this case, the resulting response functions yielded shorter lags, but the effect is highly dependent on wavelength (larger decreases in lag occurring at shorter wavelengths). It also depends on the average volume filling factor (more clouds are shadowed for higher volume filling factors).  However, for high values of the average volume filling factor, this approximation overestimates the contribution of shadowed clouds on the response function. 

TORMAC can also simulate IR light curves in response to an arbitrarily defined input UV/optical light curve, including resampled observed light curves. As an example, we have presented simulated light curves at 3.6, 4.5, and 30 $\mu$m for an input signal derived from the observed optical light curve of NGC 6418. The lags recovered from cross-correlation analysis reflect the trends seen in the response functions. In particular, they increase at shorter wavelengths when cloud orientation was included, and decrease significantly at all wavelengths when anisotropic illumination of the torus is included. In general, TORMAC can provide predictions for the time response of the torus IR spectrum that can be directly compared with observations, yielding constraints on the shape, size, and composition of the dust distribution surrounding an actively accreting SMBH.

The current version of TORMAC has some limitations that we plan to address in the near future. 
The next version of the code will include self-occultation, that is, wavelength-dependent attenuation of IR radiation from each cloud by intervening clouds along the observer's line of sight. Further developments will include more sophisticated treatments of dust sublimation and heating by the diffuse radiation field, and provision for a mixture of cloud properties (e.g., optical depth and dust composition).

\acknowledgments
This paper is based upon work supported by NASA under award Nos. NNX12AC68G and NNX16AF42G. R.N. acknowledges support by FONDECYT grant No. 3140436. We would like to thank Moshe Elitzur for productive discussions that resulted in improvements to the manuscript and Ying Zu for assistance in the use of his code JAVELIN. We also thank the anonymous referee for useful comments that improved the clarity of the paper.

\bibliography{paper_v2}

\end{document}